\documentclass[journal]{IEEEtran}
\usepackage[T1]{fontenc}
\usepackage{amsmath,amssymb,amsthm,bm,mathtools}
\usepackage{graphicx,booktabs,array,tabularx,multirow,longtable}
\usepackage{tikz}
\usetikzlibrary{arrows.meta,positioning,calc,fit}
\usepackage{cite}
\usepackage{float}
\usepackage{placeins}
\usepackage{needspace}
\usepackage[nocheckfootnote,nocheckfloat]{flushend}
\floatstyle{ruled}
\newfloat{algorithm}{tbp}{loa}
\floatname{algorithm}{Algorithm}
\usepackage[hidelinks]{hyperref}

\newtheorem{theorem}{Theorem}

\newtheorem{proposition}{Proposition}

\graphicspath{{figures/}}
\AtBeginDocument{%
  \setlength{\parskip}{0pt plus .1pt}%
  \setlength{\abovedisplayskip}{3pt plus .5pt minus .5pt}%
  \setlength{\belowdisplayskip}{3pt plus .5pt minus .5pt}%
  \setlength{\abovedisplayshortskip}{1.5pt plus .5pt}%
  \setlength{\belowdisplayshortskip}{2pt plus .5pt minus .5pt}%
  \setlength{\textfloatsep}{9pt plus 1pt minus 2pt}%
  \setlength{\dbltextfloatsep}{7pt plus 1pt minus 1pt}%
  \setlength{\floatsep}{7pt plus 1pt minus 1pt}%
  \setlength{\dblfloatsep}{7pt plus 1pt minus 1pt}%
  \setlength{\intextsep}{7pt plus 1pt minus 1pt}%
  \setlength{\jot}{2pt}%
}
\newcommand{\CAPR}{CAPR}

\newcommand{\E}{\mathbb{E}}

\newcommand{\diag}{\operatorname{diag}}

\newcommand{\ResultGfourCapture}{83.8}
\newcommand{\ResultGfourPFA}{4.68}
\newcommand{\ResultGfourEstimate}{87.8}
\newcommand{\ResultBestGain}{12.5}
\newcommand{\ResultBestGainLow}{12.3}
\newcommand{\ResultBestGainHigh}{12.8}

\newcommand{\ResultIndependentGain}{0.157}
\newcommand{\ResultIndependentGainLow}{0.0809}
\newcommand{\ResultIndependentGainHigh}{0.234}

\newcommand{\ResultGfourDelay}{0.430}
\newcommand{\ResultGfourCFO}{611}
\newcommand{\ResultDelayReduction}{4.38}
\newcommand{\ResultCFOReduction}{7.33}

\newcommand{\ResultPhysicalCapture}{83.8}
\newcommand{\ResultPooledDelay}{0.618}
\newcommand{\ResultPooledCFO}{878}
\newcommand{\ResultPooledDelayUs}{2.57}
\newcommand{\ResultMacroDelayUs}{1.78}
\newcommand{\ResultBestWins}{78}
\newcommand{\ResultBestTies}{11}
\newcommand{\ResultBestLosses}{61}

\newcommand{\ResultTests}{194560}
\newcommand{\ResultCalibration}{194560}
\newcommand{\ResultPrimaryCases}{150}

\newcommand{\ResultPeakGain}{1.05}
\newcommand{\ResultPeakLow}{0.773}
\newcommand{\ResultPeakHigh}{1.32}
\newcommand{\ResultZeroGain}{0.735}
\newcommand{\ResultZeroLow}{0.608}
\newcommand{\ResultZeroHigh}{0.863}

\newcommand{\ResultCorrReplayZero}{4.22}
\newcommand{\ResultCorrReplayEight}{3.91}
\newcommand{\ResultCorrReplayNineFive}{2.19}

\newcommand{\ResultSevereFiveCapture}{6.41}
\newcommand{\ResultSevereFiveEstimate}{32.8}

\newcommand{\ResultSevereTenCapture}{4.53}
\newcommand{\ResultSevereTenEstimate}{30.5}

\newcommand{\ResultIdentityEstimate}{99.1}
\newcommand{\ResultIdentityDetection}{6.09}
\newcommand{\ResultIdentityPFA}{5.94}
\newcommand{\ResultNearReplayCapture}{0.00}

\newcommand{\ResultNearReplayDetection}{5.16}
\newcommand{\ResultNearReplayPFA}{3.28}

\newcommand{\ResultValidActions}{193278}
\newcommand{\ResultEmptyActions}{1282}
\newcommand{\ResultMaximumGap}{0.0200}
\newcommand{\ResultMaximumOmitted}{0.372}

\newcommand{\ResultReplayCases}{32}

\newcommand{\ResultReplayGain}{54.2}

\newcommand{\ResultReplayShare}{92.1}

\newcommand{\ResultReplayCapture}{73.4}
\newcommand{\ResultReplayGaussian}{19.2}
\newcommand{\ResultOtherCases}{118}
\newcommand{\ResultOtherGain}{1.25}

\newcommand{\ResultDiffuseCapture}{90.2}
\newcommand{\ResultDiffuseGaussian}{94.4}
\newcommand{\ResultDiffuseBest}{97.3}
\newcommand{\ResultClockOODCapture}{82.0}
\newcommand{\ResultClockOODBest}{89.8}
\newcommand{\ResultEstimateGain}{1.65}

\newcommand{\ResultRejectedCorrectReduction}{10.9}

\newcommand{\ResultAcceptedIncorrectIncrease}{1.29}

\title{Capture-Aware Predictive OFDM Reacquisition in Clock-Asynchronous Space-Air-Ground Networks}

\author{Zhihan~Zeng,~\IEEEmembership{Graduate~Student~Member,~IEEE,}
        Kaihe~Wang,~\IEEEmembership{Graduate~Student~Member,~IEEE,}\\
        Zhongpei~Zhang,
        Chongwen~Huang,~\IEEEmembership{Senior~Member,~IEEE}%
\thanks{Zhihan Zeng is with the National Key Laboratory of Wireless Communications, University of Electronic Science and Technology of China (UESTC), Chengdu, China (e-mail: zzhihan@std.uestc.edu.cn).}%
\thanks{Kaihe Wang and Zhongpei Zhang are with the Shenzhen Institute for Advanced Study, UESTC, Shenzhen 518110, China (e-mail: khewang@yeah.net; zhangzp@uestc.edu.cn).}%
\thanks{Chongwen Huang is with the College of Information Science and Electronic Engineering, Zhejiang University, Hangzhou 310027, China, and also with the Zhejiang Provincial Key Laboratory of Multi-Modal Communication Networks and Intelligent Information Processing and the National Key Laboratory of Millimeter-Wave and Terahertz Remote Sensing, Hangzhou 310027, China (e-mail: chongwenhuang@zju.edu.cn).}%
\thanks{Corresponding author: Zhongpei Zhang.}}

\begin{document}
\maketitle
\bstctlcite{capr:bibcontrol}
\begin{abstract}
Reliable orthogonal frequency-division multiplexing (OFDM) reacquisition must validate an intended burst and return its delay and carrier frequency offset (CFO) corrections within capture tolerances. In clock-asynchronous space-air-ground integrated networks (SAGINs), shared geometry and timing uncertainty couples distributed observations, while public-preamble replay creates competing peaks. We propose Capture-Aware Predictive Reacquisition (CAPR), a Bayesian receiver that models competition under both hypotheses and marginalizes unknown spatial interference covariance and target complex amplitude before shared-state fusion. CAPR maximizes posterior probability mass within the capture rectangle and constrains continuous L-BFGS-B refinement through a mass-loss budget. We derive closed-form predictive likelihood reductions, an exact finite-cell action optimizer, a conditional approximation regret bound, and finite-sample false-acceptance calibration. Across \ResultPrimaryCases{} primary scenarios and five seeds, CAPR achieves \ResultGfourCapture\% strict capture at \ResultGfourPFA\% realized false acceptance, improving over Gaussian joint-MAP by \ResultBestGain{} percentage points overall and \ResultReplayGain{} points across all \ResultReplayCases{} replay scenarios. Diffuse multipath, degraded assistance, and indistinguishable sources delineate its operational boundaries.
\end{abstract}

\begin{IEEEkeywords}
OFDM synchronization, predictive inference, joint detection and estimation, satellite communications, clock uncertainty.
\end{IEEEkeywords}
\Needspace{5\baselineskip}
\section{Introduction}
\label{sec:introduction}
\IEEEPARstart{S}{pace}-air-ground integrated networks (SAGINs) unite the wide coverage of low-Earth-orbit (LEO) satellites, the agile deployment of unmanned aerial vehicles (UAVs), and the high capacity of terrestrial base stations (BSs)~\cite{liu2018sagin,kodheli2021satellite,zeng2016uav}. Receiving an uplink burst across such heterogeneous nodes provides valuable macro-diversity, yet distinct platform altitudes and trajectories induce disparate propagation delays, Doppler shifts, and clock drifts. Large Doppler excursions dominate LEO links~\cite{you2020leomimo}, whereas air-to-ground links depend strongly on elevation geometry and local scattering~\cite{khawaja2019airground,huang2022holographic_mimo_channel_modeling,xu2022twotier_ris_noma_zte,shen2024nearfield_beam_training_holographic_mimo,wei2023ris_model_based_model_free_jii,jiang2024ris_noma_relay_jii,xu2023reconfiguring_wireless_environments_scis}. These physical discrepancies become tightly coupled when distributed nodes must collaboratively reacquire an interrupted connection before a backhaul delivery deadline.

Consider a previously served terminal transmitting a known public OFDM preamble to restore uplink synchronization after a brief outage. The network retains coarse prior estimates of terminal position and relative node clocks, both of which degrade with age. Participating BS, UAV, and satellite arrays collect target-silent reference windows alongside the acquisition window and forward quantized in-phase and quadrature (IQ) packets to a fusion processor, where only timely arrivals contribute to the decision. Meanwhile, an uncoordinated user or replayer transmitting the same public preamble can create a strong competing peak. The receiver cannot merely detect preamble energy; it must determine whether the intended burst is present and return delay and CFO corrections that fall inside the demodulator's capture tolerances.

Classical OFDM acquisition exploits repeated symbols for CFO estimation~\cite{moose1994frequency} and structured training for joint timing and frequency recovery~\cite{schmidl1997robust}, with multiuser extensions handling unequal offsets~\cite{morelli2007ofdma}. Under low signal-to-noise ratio (SNR) or high mobility, prior assistance helps narrow the search space: timing assistance restricts LEO delay-Doppler grids~\cite{li2026tiad}, while bidirectional frequency exchanges separate kinematics from clock drift in UAV networks~\cite{jin2024fidtml}. In distributed reception, Bayesian localization and synchronization jointly model geometry and clocks in MIMO-OFDM~\cite{teng2026blacs}, and feedback resampling aids timing recovery across non-homologous satellite receivers~\cite{zhang2026timing}. For single-burst reacquisition, network assistance specifies a distribution over a shared latent state: a single terminal-position error perturbs delay, Doppler, and angle of arrival (AoA) across all arrays, while timing exchanges couple their relative clocks. Treating node uncertainties independently can therefore assign false credibility to local peaks that are incompatible with any common terminal state, whereas fusing weak observations through a shared state reinforces consistent evidence.

Structured interference introduces a distinct ambiguity. Pilot reuse is well known to contaminate spatial channel estimation~\cite{marzetta2010massive,jose2011pilot}. Interference-resilient synchronization exploits adjacent-channel structure~\cite{grotz2007joint}, randomized probing~\cite{darsena2022antijamming}, rank-one projection~\cite{marti2025jass}, or satellite-swarm coordination~\cite{beidas2026swarm}, while related statistical methods employ covariance-based activity detection~\cite{fengler2021nonbayesian,li2023asynccov}, sparse delay-Doppler recovery~\cite{zuo2023ofdm}, and learned covariance features against pilot spoofing~\cite{choudhury2024spoofing}. However, spatial covariance captures only cross-antenna correlation, whereas a competing preamble also exhibits coherent temporal structure across the burst. Moreover, guarantees predicated on secret synchronization sequences~\cite{marti2025jass} do not apply to acquisition using a public preamble.

Together, these observations expose two coupled difficulties. First, coherent hypothesis testing under shared uncertainty requires representing the competing preamble under both the null ($\mathcal{H}_0$) and alternative ($\mathcal{H}_1$) hypotheses; otherwise, structured interference energy is easily misattributed to the target. Because short reference windows leave the spatial covariance uncertain and the target complex amplitude is unknown, the receiver must marginalize over both nuisances while propagating shared geometry and clock priors across hypotheses. Second, downstream demodulation succeeds only if the returned correction lands inside a prescribed capture rectangle. Because a sharp posterior mode can contain less probability mass within this tolerance box than a broader neighboring region, maximum \emph{a posteriori} (MAP) point estimation need not maximize capture probability. While joint detection and estimation theory links acceptance to estimation quality~\cite{moustakides2012joint}, practical implementation requires an evaluable predictive distribution and an acceptance score tied to the returned action, so that any capture-mass loss caused by continuous likelihood refinement is explicitly controlled.

To address these challenges, we propose Capture-Aware Predictive Reacquisition (CAPR). Rather than substituting plug-in point estimates, CAPR embeds a common competing-preamble mixture under both hypotheses and analytically integrates out spatial covariance and target complex amplitude. Projecting whitened observations onto target-aligned coordinates reduces the complex amplitude integral to a scalar Gamma expectation, while null-posterior reweighting compares target evidence directly against competing explanations. Local conditional likelihoods are fused across nodes separately under each hypothesis before integrating over the shared assistance state, preserving geometric and clock coupling. On the decision side, CAPR scores actions by their posterior capture mass, finds the exact cell-posterior optimum via finite breakpoint enumeration, and gates continuous L-BFGS-B profile refinement with an explicit mass-loss budget whose resulting mass updates the final log-detection score.

The main contributions are as follows:

1) We derive a normalized predictive likelihood that integrates spatial covariance and target complex amplitude alongside a common competitor and shared assistance state, establishing exact rank-reduced determinant updates and distinguishing the complete probability law from its screened numerical implementation.

2) We formulate capture-mass action selection, prove an exact finite-breakpoint optimizer, and constrain continuous refinement via the returned action's mass. We establish normalized-model capture optimality, a conditional regret bound under total variation (TV) posterior discrepancy, shared-state evidence shift identities, and finite-sample calibration via exchangeable null scores.

3) We benchmark CAPR against six baselines and three component ablations across \ResultPrimaryCases{} primary scenarios, two negative controls, and five seeds (389,120 calibration and test observations). Jointly evaluating strict capture, false acceptance, threshold-free estimation, and root-mean-square error (RMSE) separates estimation accuracy from acceptance calibration.

Our evaluation incorporates unmodeled physical stresses including multipath echoes, fractional sampling mismatch, finite-bit quantization, and backhaul deadlines. Across the benchmark, CAPR improves mean strict capture by \ResultBestGain{} percentage points (pp) over Gaussian joint-MAP, with \ResultReplayShare\% of the aggregate gain concentrated in the \ResultReplayCases{} replay scenarios (\ResultReplayCapture\% versus \ResultReplayGaussian\%). We also document performance boundaries under severe low SNR, diffuse channels, clock-prior mismatch, and indistinguishable replay.

Section~\ref{sec:system} formulates the system model and reacquisition objective. Section~\ref{sec:predictive} develops the CAPR receiver. Section~\ref{sec:analysis} presents the theoretical analysis. Section~\ref{sec:setup} describes the simulation setup, Section~\ref{sec:experiments} reports numerical results and operational boundaries, and Section~\ref{sec:conclusion} concludes the paper. Derivations and proofs are given in Appendices~\ref{app:predictive} and~\ref{app:decisionproofs}.

\section{System Model and Problem Formulation}
\label{sec:system}
\subsection{Network and Signal Model}
\begin{figure}[t]
\centering
\input{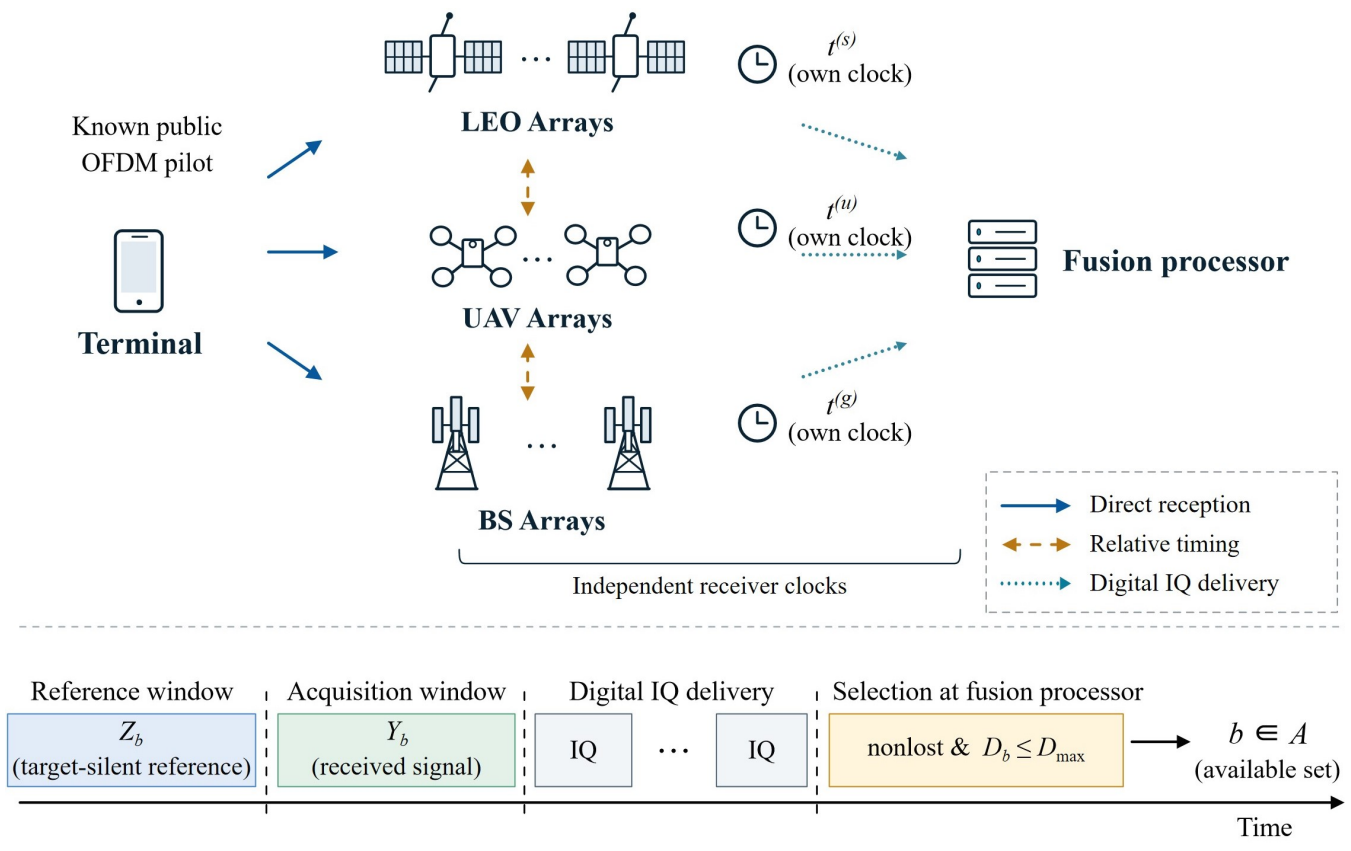}
\caption{Direct reception, relative timing, and digital IQ delivery. Solid, dashed, and dotted arrows distinguish these links. Target-silent reference samples inform nuisance covariance; the deadline determines available observations.}
\label{fig:system}
\end{figure}

A terminal sends a public OFDM burst to $B$ BS, UAV, and LEO arrays indexed by $\mathcal B$. Each $M_b$-port node exchanges relative timing and forwards digital IQ to reference node $0$ (Fig.~\ref{fig:system}). Reception is coherent within each array and burst; independent complex amplitudes permit noncoherent fusion across nodes.

In an Earth-centered inertial frame, terminal/node positions $\bm p(t),\bm r_b(t)$ define $\rho_b=\|\bm r_b-\bm p\|$, $\tau_b=\rho_b/c$, and $\nu_{D,b}=-f_c\dot\rho_b/c$. Trajectories combine circular LEO orbits, Earth rotation, and UAV motion; nominal compensation uses simultaneous-state ranges without retarded-time or relativistic corrections. At epoch $t_e$, clock $C_b(t_e+u)-t_e=(1+\epsilon_b)u+\beta_b$ gives first-order inverse time error $-\beta_b-\epsilon_bu$ and CFO $-f_c\epsilon_b$. The target state is $\bm x=(d,f)^T$, with delay $d=f_s\tau$ in samples and normalized CFO $f=\nu/\Delta\nu$; $\Delta\nu$ differs from subcarrier spacing. Relative clocks $d_b^c=f_s\beta_b$, $f_b^c=-f_c\epsilon_b/\Delta\nu$ are anchored at node $0$. Shared position/velocity error $\bm g=(\delta\bm p_0^T,\delta\bm v_0^T)^T$ has six coordinates, or four with zero vertical errors; platforms remain three-dimensional. Stacking clocks and geometry into $\bm\eta$ gives
\begin{equation}\label{eq:local-geometry-maps}
\begin{aligned}
d_b(\bm x,\bm\eta)&=d+d_b^c+J_{d,b}\bm g,\\
f_b(\bm x,\bm\eta)&=f+f_b^c+J_{f,b}\bm g,\\
u_b(\bm\eta)&=u_{b,0}+J_{u,b}\bm g.
\end{aligned}
\end{equation}
Direction cosine $u_b$ is clipped to $[-1,1]$ before steering. Set $\bm e_b=(\bm r_b-\bm p)/\rho_b$, $\bm v_b^{\rm rel}=\dot{\bm r}_b-\dot{\bm p}$, $G=\partial\bm p/\partial\bm g$, and $\dot G=\partial\dot{\bm p}/\partial\bm g$. Then $\dot\rho_b=\bm e_b^T\bm v_b^{\rm rel}$ and
\begin{equation}\label{eq:geometry-chain}
\begin{aligned}
\partial_{\bm g}\rho_b&=-\bm e_b^TG,\\
\partial_{\bm g}\bm e_b&=-\rho_b^{-1}(I-\bm e_b\bm e_b^T)G,\\
\partial_{\bm g}\dot\rho_b
 &=-\rho_b^{-1}(\bm v_b^{\rm rel})^T(I-\bm e_b\bm e_b^T)G
   -\bm e_b^T\dot G.
\end{aligned}
\end{equation}
For six coordinates, $G(t)=R_E(t)[I,tI]$ rotates reference axes into the inertial frame; $\dot G$ includes Earth rotation and velocity propagation. Four coordinates restrict these columns. Thus $J_{d,b}=(f_s/c)\partial_{\bm g}\rho_b$, $J_{f,b}=-f_c(c\Delta\nu)^{-1}\partial_{\bm g}\dot\rho_b$, and $J_{u,b}=\rho_b^{-1}(\bm a_b^{\rm ax})^T(I-\bm e_b\bm e_b^T)G$ for array axis $\bm a_b^{\rm ax}$, since $u_b=-(\bm a_b^{\rm ax})^T\bm e_b$. Geometry therefore couples delay, Doppler, and direction across nodes. Inference uses centered finite differences at the reference epoch; generation propagates perturbed trajectories across the burst, leaving finite-difference and linearization mismatch.

With $N_s$ OFDM symbols, FFT size $N_F$, prefix $N_{\rm cp}$, and two $N_g$-sample guards, pilot length is $N=N_s(N_F+N_{\rm cp})+2N_g$. Its finite DFT interpolant is $p_c(v)=\sum_{k\in\mathcal K}\widetilde p_ke^{\mathrm i2\pi kv/N}$, with signed indices $\mathcal K$ and inverse-DFT scaling in $\widetilde p_k$. Set $t_n=[n-(N-1)/2]/f_s$ and $\Delta\rho_b(u)=\rho_b(t_e+u;\bm g)-\rho_b(t_e+u;\bm0)$. The physical template is
\begin{equation}\label{eq:physical-waveform}
\begin{aligned}
v_{b,n}&=n-d-d_b^c-\epsilon_b[n-(N-1)/2]
            -f_s\Delta\rho_b(t_n)/c,\\
\phi_{b,n}&=2\pi(f+f_b^c)\Delta\nu t_n
 -\tfrac{2\pi f_c}{c}[\Delta\rho_b(t_n)-\Delta\rho_b(0)],\\
s_b[n]&=p_c(v_{b,n})e^{\mathrm i\phi_{b,n}}.
\end{aligned}
\end{equation}
Channel amplitude absorbs constant range phase; differential phase already includes residual Doppler. Skew and CFO precede Fourier processing, preserving intercarrier interference (ICI); guards limit interpolant wraparound. For a block with centered origin $n_c$, put $d_{\rm blk}=d-\epsilon n_c$ and absorb constant carrier phase into amplitude. Constant skew $\epsilon$ and CFO $\nu$ yield
\begin{equation}\label{eq:ici-kernel}
 C_{k\ell}=\frac1{N_F}\sum_{n=0}^{N_F-1}
 e^{\mathrm i2\pi[(\ell-k-\ell\epsilon+N_F\nu/f_s)n-\ell d_{\rm blk}]/N_F}.
\end{equation}
Off-diagonal terms are ICI. Joint timing/frequency estimation is classical~\cite{vandeBeek1997ml}; \eqref{eq:physical-waveform} also retains time-varying range residuals.

Under $H_1$, the $M_b\times N$ acquisition matrix is
\begin{equation}\label{eq:physical-observation}
Y_b=\bm h_bs_b^T+\sum_\ell\bm h_{b,\ell}s_{b,\ell}^T+J_b+U_b+W_b.
\end{equation}
The direct channel is Rician; remaining terms are optional echoes, interference $J_b$, extra pilots $U_b$, and unit-variance circular Gaussian noise $W_b$. Under $H_0$, target and echoes are absent. Interference covers common/node-local public-preamble replay, tones, chirps, barrage, and two spatial noise components. Inference represents covariance, coherent competition, amplitude/direction, and assistance uncertainty; scattering, arbitrary echoes, phase noise, impulses, and extra pilots test mismatch.

Each node supplies target-silent $Z_b\in\mathbb C^{M_b\times T_b}$. Nuisance spatial channels persist across reference/capture with independent temporal samples and varying reference visibility. Reference/capture IQ and a diagnostic $N_F$-sample payload block share a peak-component quantization scale. At $q_b>0$ bits per real component, $L_b=2M_b(N+N_F+T_b)q_b+32$ bits; float delivery uses $64M_b(N+N_F+T_b)$ bits. Nominal uplink/backhaul ranges $\rho_{b,0},\rho_{b\to0}$, rate $R_b$, and exponential queue delay $Q_b$ give
\begin{equation}\label{eq:delivery}
D_b=(\rho_{b,0}+\rho_{b\to0})/c+(N+N_F+T_b)/f_s+L_b/R_b+Q_b.
\end{equation}
Nonlost packets meeting $D_b\le D_{\max}$ define the common delivered set $\mathcal A$; an empty set is rejected with a zero placeholder estimate. Controlled-SNR cases adjust transmit power to a target median pre-shadowing link SNR; fixed-power cases retain the physical link budget. Evaluation covers synchronization capture and link delivery delay; receiver CPU profiling and payload decoding lie outside this benchmark.

\subsection{Problem Formulation}
\label{sec:problem}
For action $\bm a=(a_d,a_f)^T$, define
\begin{equation}\label{eq:capture-region}
\mathcal C(\bm a)=\{\bm x:|d-a_d|\le r_d,\ |f-a_f|\le r_f\}.
\end{equation}
Default tolerances are $r_d=0.5$ sample and $r_f=750\,\mathrm{Hz}/\Delta\nu=0.5$. Given assistance $\mathcal I$, let $\delta(Y)\in[0,1]$ be the acceptance probability, $P_{\rm A}=\E_1[\delta(Y)\mathbf1\{\bm x\in\mathcal C(\bm a(Y))\}\mid\mathcal I]$ the strict capture probability, and $P_{\rm FA}=\E_0[\delta(Y)\mid\mathcal I]$ the false-acceptance rate. The objective is
\begin{equation}\label{eq:task-risk}
\max_{\delta,\bm a}P_{\rm A}\qquad\text{subject to }P_{\rm FA}\le\alpha_{\rm FA}.
\end{equation}
Rejecting a correct estimate or accepting an out-of-tolerance estimate fails this task. The uniform working prior on $\mathcal X=[-3,3]^2$ uses square cells; physical test offsets are continuous on $[-2.2,2.2]^2$. A companion metric fixes delay tolerance at $0.5/240000$ s, or $2.08\,\mu$s, across sample rates.

Relative clocks form $\bm q=[d_1^c,\ldots,d_{B-1}^c,f_1^c,\ldots,f_{B-1}^c]^T$. For a connected timing graph~\cite{freris2011clocks,simeone2008distributed}, the incidence matrix $D$ omits the reference column; oriented differences satisfy $\bm z=H\bm q+\bm e$, $H=\diag(D,D)$. Bidirectional frequency sums and differences separate radial motion from relative clock CFO~\cite{jin2024fidtml}; the graph uses the resulting sufficient statistics. With $k_c=-f_s\Delta\nu/f_c$, a clock pair obeys $\mathrm d d^c=k_cf^c\,\mathrm dt$ and $\mathrm d f^c=\sqrt{q_c}\,\mathrm dW$. Integrating Brownian increments over assistance age $a$ gives
\begin{equation}\label{eq:clock-age}
\begin{split}
P_q(a)&=\left(F_aP_{\rm pair}F_a^T+
 q_c\begin{bmatrix}k_c^2a^3/3&k_ca^2/2\\k_ca^2/2&a\end{bmatrix}\right)\otimes C_\rho,\\
F_a&=\begin{bmatrix}1&k_ca\\0&1\end{bmatrix},\qquad
C_\rho=(1-\rho)I+\rho\bm1\bm1^T.
\end{split}
\end{equation}
Here $P_{\rm pair}=\diag(0.45^2,0.85^2)$ and $q_c=0.035^2$. The covariance is positive definite for evaluated $\rho$; the cross-term sign follows $k_c<0$.

For row $\bm h_e^T$ of $H$ and measurement standard deviation $\sigma_e$, residuals $r_e=(z_e-\bm h_e^T\bm q)/\sigma_e$ define the Student-$t$ objective $J(\bm q)=\bm q^TP_q^{-1}\bm q/2+\sum_e(\nu_g+1)\log(1+r_e^2/\nu_g)/2$. Seven weighted solves start with unit weights and use $\nu_g=4$:
\begin{equation}\label{eq:graph-update}
\begin{aligned}
K&=P_q^{-1}+H^T\diag(w_e/\sigma_e^2)H,\\
\widehat{\bm q}&=K^{-1}H^T\diag(w_e/\sigma_e^2)\bm z,\quad
w_e\leftarrow(\nu_g+1)/(\nu_g+r_e^2).
\end{aligned}
\end{equation}
After the final weight update, $\widetilde P_q=K^{-1}$ is paired with the seventh mean, defining a finite-budget Gaussian surrogate, not the exact graph posterior. Independent clock and geometry blocks, with diagonal reference-epoch geometry covariance $P_g$, give
\begin{equation}\label{eq:assistance-distribution}
q(\bm\eta\mid\mathcal I)=\mathcal N\!\left(
\begin{bmatrix}\widehat{\bm q}\\\bm0\end{bmatrix},
\begin{bmatrix}\widetilde P_q&0\\0&P_g\end{bmatrix}\right).
\end{equation}
The shared state couples local parameters through \eqref{eq:local-geometry-maps}, as in waveform-based geometric inference~\cite{shen2010localization}. Given $\bm x,\bm\eta,Z_b$, node-specific nuisances are independent: local densities multiply before the common-state integral. Independent state copies define an ablation; graph and reference evidence each enter once. Availability conditions on nominal-range delivery, queueing, and loss; state-dependent selection requires a conditioned assistance model. Data-dependent quantization remains a physical mismatch.

\section{Proposed Method}
\label{sec:predictive}\label{sec:receiver}
Reference-trained processing underlies adaptive detection~\cite{reed1974rapid,kelly1986adaptive,robey1992cfar}; subspace and robust covariance methods address structured or non-Gaussian interference~\cite{scharf1994matched,conte2002recursive,wiesel2012geodesic}. \CAPR{} integrates covariance and amplitude within a common competitor mixture under both hypotheses. We condition on assistance $\mathcal I$, reference blocks, and delivered set $\mathcal A$, suppressing local node indices.

\subsection{Predictive Observation Model}
Let $Y\in\mathbb C^{M\times N}$ and $Z\in\mathbb C^{M\times T}$ denote capture and reference blocks. We use column vectorization; $T$, $H$, and $*$ denote transpose, conjugate transpose, and conjugation, with $\mathcal{CN}(z;m,C)=e^{-(z-m)^HC^{-1}(z-m)}/(\pi^n|C|)$. For Hermitian $\Psi\succ0$ and $\kappa>M-1$, the complex inverse-Wishart density is
\begin{equation}
 p(\Sigma\mid\Psi,\kappa)=
 \frac{|\Psi|^\kappa}{\Gamma_M(\kappa)}
 |\Sigma|^{-(\kappa+M)}e^{-\operatorname{tr}(\Psi\Sigma^{-1})},
 \label{eq:pred-ciw}
\end{equation}
where $\Sigma\succ0$ is $M\times M$ and $\Gamma_M(a)=\pi^{M(M-1)/2}\prod_{m=1}^M\Gamma(a-m+1)$. Prior $\operatorname{CIW}_M(I_M,M+2)$ and conditionally independent reference columns $\mathcal{CN}(0,\Sigma)$ give
\begin{equation}
 \Sigma\mid Z\sim\operatorname{CIW}_M(\Psi,\kappa),
 \label{eq:pred-reference}
\end{equation}
with $\Psi=I_M+ZZ^H$ and $\kappa=M+2+T$, proper even at $T=0$. Reference coherence, interference changes, and quantization test this independence approximation. For $A\in\mathbb C^{M\times N}$, $V\succ0$ of size $N\times N$, and $E=Y-A$, assume $\operatorname{vec}(E)\mid\Sigma\sim\mathcal{CN}(0,V\otimes\Sigma)$:
\begin{equation}
 p(Y\mid A,\Sigma,V)=
 \frac{e^{-\operatorname{tr}(\Sigma^{-1}EV^{-T}E^H)}}
 {\pi^{MN}|V|^M|\Sigma|^N}.
 \label{eq:pred-matrix-normal}
\end{equation}
Here $V^{-T}$ follows from column vectorization with $A=\alpha\bm a_{\rm s}\bm s^T$. Marginalizing $\Sigma$ yields
\begin{equation}
 p(Y\mid A,V,Z)=C_Z|V|^{-M}|\Psi+EV^{-T}E^H|^{-K},
 \label{eq:pred-matrix-student}
\end{equation}
where $K=\kappa+N$ and $C_Z=\Gamma_M(K)|\Psi|^\kappa/[\pi^{MN}\Gamma_M(\kappa)]$. Only $C_Z$ is candidate/component independent. Appendix~\ref{app:predictive} proves normalization and the amplitude reduction.

For unit-energy competitor $\bm v_j$ and its delay/CFO derivatives, form $T_j=[\bm v_j,\ h_J\bm v_{d,j}/\sqrt{12},\ h_J\bm v_{f,j}/\sqrt{12}]$ and
\begin{equation}
 V_{j,l}=I_N+g_lT_jT_j^H.
 \label{eq:pred-tangent-cov}
\end{equation}
Independent centered uniform offsets have variance $h_J^2/12$, so $T_jT_j^H$ matches the first-order waveform's second moment, with remainder norm at most $L_j\|\bm\delta\|^2/2$ under Hessian bound $L_j$. This Gaussian moment approximation does not specify the full displaced-waveform law. Competitor spatial coefficients have covariance proportional to $\Sigma$; the unit-norm target direction $\bm a_{\rm s}(\bm\eta)$ remains explicit.

The grid uses spacing $h_J=0.5$, radius $\max(12,R_{\rm loc})$ with local target lookup radius $R_{\rm loc}=6$, $D_J$ waveform points, and $P_J$ powers. The no-competitor component has $V_0=I_N$, $\pi_0=1-\omega$; each waveform/power pair has mass $\omega/(D_JP_J)$. Defaults are $\omega=0.20$ and $g_l=E_p10^{b_l/10}$, $b_l\in\{-30,-20,-10,0,10,20,30\}$ dB, and unnormalized preamble energy $E_p$. Let $j$ index each pair and $p_{0,j}(Y)=p(Y\mid0,V_j,Z)$. For unit-energy target $\bm s(\bm x,\bm\eta)$,
\begin{align}
 p_0(Y)&=\sum_j\pi_jp_{0,j}(Y),
 \label{eq:pred-null-mixture}\\
 p_T(Y\mid\bm x,\bm\eta)
 &=\sum_{j,p}\pi_jw_p\int_{\mathbb C}
 p(Y\mid\alpha\bm a_{\rm s}\bm s^T,V_j,Z)\nonumber\\
 &\hspace{13mm}\times\mathcal{CN}(\alpha;0,\gamma_p)\,\mathrm d^2\alpha.
 \label{eq:pred-target-mixture}
\end{align}
where $w_p=1/P_T$ and $\gamma_p=ME_p10^{a_p/10}$ for $a_p\in\{-30,-22,-14,-6,2\}$ dB; $M$ accounts for the unit-norm spatial response. These complete fixed mixtures are normalized.

\subsection{Marginal Likelihood Reduction}
Diagonalize $T_jT_j^H=U_j\operatorname{diag}(\lambda_{j\ell})U_j^H$ with $r_j\le3$ orthonormal columns. Set $D_j=\operatorname{diag}\{g_j\lambda_{j\ell}/(1+g_j\lambda_{j\ell})\}$, $C_j=YU_j^*$, and $R=\Psi+YY^H$. Then $V_j^{-1}=I_N-U_jD_jU_j^H$ and
\begin{equation}
 R_j=\Psi+YV_j^{-T}Y^H=R-C_jD_jC_j^H.
 \label{eq:pred-lowrank-null}
\end{equation}
reducing each null determinant update to at most $3\times3$; Appendix~\ref{app:predictive} supplies the inverse identity. For each target template, define
\begingroup\interdisplaylinepenalty=10000
\begin{align}
 e_j&=\bm s^HV_j^{-1}\bm s,&
 \bm c_j&=Y(V_j^{-1}\bm s)^*/\sqrt{e_j},\nonumber\\
 S_j&=R_j-\bm c_j\bm c_j^H,&q_j&=\bm c_j^HS_j^{-1}\bm c_j,\nonumber\\
 u_j&=\bm a_{\rm s}^HS_j^{-1}\bm a_{\rm s},&
 v_j&=|\bm a_{\rm s}^HS_j^{-1}\bm c_j|^2.
 \label{eq:pred-target-projections}
\end{align}
\endgroup
Whitening and rotation isolate $\bm c_j$ from $N-1$ target-free columns $Y_\perp$, giving $S_j=\Psi+Y_\perp Y_\perp^H\succ0$ and $e_j,u_j>0$. Conditional on $Y_\perp$, introduce $z\sim\operatorname{Gamma}(K-M,1)$ in shape--rate convention:
\begin{equation}
 \bm c_j\mid z,H_0\sim\mathcal{CN}(0,S_j/z).
 \label{eq:pred-gamma-prior}
\end{equation}
Observing $\bm c_j$ contributes $z^Me^{-zq_j}$, yielding null posterior $\operatorname{Gamma}(K,r)$ with $r=1+q_j$. At fixed $z$, integrating $\alpha\sim\mathcal{CN}(0,\gamma_p)$ adds covariance $\rho\bm a_{\rm s}\bm a_{\rm s}^H$ to $S_j/z$, where $\rho=\gamma_pe_j$, and gives
\begin{equation}
 B_{T\mid j,p}=\mathbb E_{\operatorname{Gamma}(K,r)}
 \frac{\exp\{\rho v_jz^2/(1+\rho u_jz)\}}{1+\rho u_jz}.
 \label{eq:pred-gamma-bf}
\end{equation}
Thus the expectation uses the null posterior and retains whitened energy $e_j$ despite $\|\bm s\|=1$. Cauchy--Schwarz gives $\theta=v_j/u_j\le q_j$, hence $r-\theta\ge1$. Exponential tilting yields
\begin{equation}
 B_{T\mid j,p}=\left(\frac r{r-\theta}\right)^K
 \mathbb E_{\operatorname{Gamma}(K,r-\theta)}
 \frac{e^{-\theta z/(1+\rho u_jz)}}{1+\rho u_jz}.
 \label{eq:pred-tilted-bf}
\end{equation}
whose integrand lies in $(0,1]$. We use this form when $\rho u_jK/r\ge1$ and \eqref{eq:pred-gamma-bf} otherwise, with 16-node normalized generalized-Laguerre quadrature and log-sum-exp accumulation. Positive weights preserve nonnegativity without a universal finite-rule error bound.

\subsection{Hypothesis Marginalization and Evidence Fusion}
Since $p_0>0$, null evidence updates competitor weights in the exact local ratio:
\begin{align}
 w_j^{(0)}(Y)&=\frac{\pi_jp_{0,j}(Y)}{p_0(Y)},\quad\sum_jw_j^{(0)}=1,
 \label{eq:pred-null-weights}\\
 L(Y;\bm x,\bm\eta)&=\frac{p_T(Y\mid\bm x,\bm\eta)}{p_0(Y)}
 =\sum_jw_j^{(0)}\sum_pw_pB_{T\mid j,p}.
 \label{eq:pred-null-reweight}
\end{align}
Target-power weights remain within each component. Retain $j=0$ and the eight largest competitor weights without renormalization. For this data-dependent set $\mathcal J(Y)$,
\begin{equation}
 \widetilde p_T=\sum_{j\in\mathcal J(Y)}\pi_jp_{0,j}\sum_pw_pB_{T\mid j,p}\le p_T.
 \label{eq:pred-screened}
\end{equation}
so $\int\widetilde p_T\,\mathrm dY\le1$, while $p_0$ remains complete. Omitted null mass cannot bound omitted target evidence; quadrature and floating-point safeguards add further approximations.

Arithmetic interpolation $L_{\rm int}=\sum_c\lambda_cL_c$, $\lambda_c\ge0$, $\sum_c\lambda_c=1$, preserves density normalization for complete inputs and weights independent of capture IQ, or the subdensity property for screened inputs. Out-of-range delay/CFO lookups use $L=1$ with the full common null; angular coordinates are clipped before interpolation.

Symmetric contamination, $0<\varepsilon<1/2$, defines
\begin{equation}
 p_{1,b}^{\varepsilon}=(1-\varepsilon)p_{T,b}+\varepsilon p_{0,b},\quad
 p_{0,b}^{\varepsilon}=(1-\varepsilon)p_{0,b}+\varepsilon p_{T,b}.
 \label{eq:pred-contamination}
\end{equation}
For $\ell_b=\log L_b$, the conditional log ratio obeys
\begin{equation}
 \psi_\varepsilon(\ell_b)=
 \log\frac{(1-\varepsilon)e^{\ell_b}+\varepsilon}
 {(1-\varepsilon)+\varepsilon e^{\ell_b}},\quad
 |\psi_\varepsilon|\le\log\frac{1-\varepsilon}{\varepsilon}.
 \label{eq:pred-contaminated-logratio}
\end{equation}
Setting $\varepsilon=0$ removes this finite bound. Contamination makes the null candidate-dependent, so both hypotheses integrate the normalized candidate prior $\pi_x$:
\begin{align}
 m_h(Y\mid\bm x,\mathcal I)
 &=\int\prod_{b\in\mathcal A}p_{h,b}^{\varepsilon}(Y_b\mid\bm x,\bm\eta)
 q(\bm\eta\mid\mathcal I)\,\mathrm d\bm\eta,\nonumber\\
 p_h(Y\mid\mathcal I)&=\int_{\mathcal X}\pi_x(\bm x)m_h(Y\mid\bm x,\mathcal I)\,\mathrm d\bm x.
 \label{eq:pred-network-marginals}
\end{align}
Under $H_0$, $\bm x$ is auxiliary. Conditional independence and normalized mixing measures normalize these complete laws. With $r_{1,b}=(1-\varepsilon)L_b+\varepsilon$ and $r_{0,b}=(1-\varepsilon)+\varepsilon L_b$, nonnegative shared-node weights $(\bm\eta_q,\pi_q)$ summing to one give
\begin{equation}
 \ell_h(\bm x_k)=\log\sum_q\pi_q
 \exp\!\left\{\sum_{b\in\mathcal A}\log r_{h,b}(Y_b;\bm x_k,\bm\eta_q)\right\}.
 \label{eq:pred-network-logfields}
\end{equation}
The common factor $\prod_bp_{0,b}(Y_b)$ cancels between separately formed marginals; both still require candidate integration. Defaults use 64 equally weighted nodes and uniform prior on equal-area cells. Independent fusion factors receiver expectations. These complete laws and the screened numerical posterior remain distinct; the local ratio bound does not ensure robustness to arbitrary model or assistance errors.

\subsection{Bayesian Capture-Optimal Decision Rule}
Under \eqref{eq:pred-network-marginals}, define $p_1(Y,\bm x\mid\mathcal I)=\pi_x(\bm x)m_1(Y\mid\bm x,\mathcal I)$. Normalizing a screened parameter posterior does not normalize its IQ-data law. For compact action domain $\mathcal D$, let $G_\star(Y)=\max_{\bm a\in\mathcal D}G(Y,\bm a)$, where
\begin{equation}
 G(Y,\bm a)=\int_{\mathcal C(\bm a)}p_1(Y,\bm x\mid\mathcal I)\,d\bm x.
 \label{eq:decision-capture-joint}
\end{equation}
A randomized acceptance probability $\delta(Y)\in[0,1]$ gives $P_A=\int\delta(Y)G(Y,\widehat{\bm x})\,dY$ and $P_{\rm FA}=\int\delta(Y)p_0(Y\mid\mathcal I)\,dY$.
\begin{theorem}[Capture-optimal reference]\label{thm:capture-optimal}
Suppose \eqref{eq:pred-network-marginals} is normalized, $G$ is measurable in $Y$ and continuous in $\bm a$, and a measurable maximizer exists. For $0<\alpha_{\rm FA}<1$, an optimum of $\max P_A$ subject to $P_{\rm FA}\le\alpha_{\rm FA}$ returns a maximizer of $G$ and thresholds
\begin{equation}
 \Lambda_A(Y)=G_\star(Y)/p_0(Y\mid\mathcal I).
 \label{eq:capture-optimal-score}
\end{equation}
Boundary randomization supplies the required null probability. Accept freely when $p_0=0<G_\star$; decisions where both vanish are immaterial.
\end{theorem}
Appendix~\ref{app:decisionproofs} proves this specialization of joint detection--estimation theory~\cite{moustakides2012joint}. Since $G(Y,\bm a)=p_1(Y\mid\mathcal I)M(\bm a\mid Y)$, the rule weights marginal evidence by posterior capture mass. Its optimality concerns the complete normalized working model.

\subsection{Exact Finite-Cell Action Optimization}\label{sec:action}
Partition the support into cells $B_{ij}=[u_i^-,u_i^+]\times[v_j^-,v_j^+]$ with widths $h_i,g_j>0$, centers $\bm x_{ij}$, and prior masses $\pi_{ij}$. Center likelihoods give
\begin{equation}
 P_{ij}=\frac{\pi_{ij}e^{\ell_1(\bm x_{ij})}}{\sum_{k,l}\pi_{kl}e^{\ell_1(\bm x_{kl})}}.
 \label{eq:decision-cell-posterior}
\end{equation}
The cell density is $P_{ij}/(h_i g_j)$ on $B_{ij}$; a uniform continuous prior requires $\pi_{ij}\propto h_i g_j$, yielding softmax for equal cells. Its exact capture mass is
\begin{align}
 W_i(a_d)&=\frac{[\min(a_d+r_d,u_i^+)-\max(a_d-r_d,u_i^-)]_+}{h_i},\nonumber\\
 V_j(a_f)&=\frac{[\min(a_f+r_f,v_j^+)-\max(a_f-r_f,v_j^-)]_+}{g_j},\nonumber\\
 M(\bm a\mid Y)&=\bm W(a_d)^TP\bm V(a_f).
 \label{eq:decision-cell-mass}
\end{align}
Exterior mass is zero, without boundary renormalization.
\begin{proposition}[Finite action enumeration]\label{prop:cell-optimum}
For a rectangular compact action set, a maximum of \eqref{eq:decision-cell-mass} occurs at a Cartesian product of action boundaries and cell edges shifted by $\pm r_d$ or $\pm r_f$, clipped to their action intervals.
\end{proposition}
Overlaps are affine between breakpoints, making each action rectangle's mass bilinear and bounded by corner values (Appendix~\ref{app:decisionproofs}). Implementation uses $[-3,3]^2$, $24\times24$ cells of width $0.25$, and $r_d=r_f=0.5$. Plateau tie-breaking within $10^{-12}$ favors the posterior mean, which is also tested. The initial action may therefore lose $10^{-12}$ mass, while its score uses enumerated maximum $M_\star$.

\subsection{Refinement and Implementation}
At the cell center nearest the initial action, freeze alternative-posterior mean delay, CFO, and direction corrections over shared nodes, together with distortionless filters constructed from $Y_bY_b^H+10^{-6}I$ and assisted directions. For projected data $\bm z_b$, profiling amplitude $\widehat\alpha_b=\bm s_b^H\bm z_b/\|\bm s_b\|^2$ gives
\begin{equation}
 J_r(\bm x)=\frac1{|\mathcal A|}\sum_{b\in\mathcal A}
 \frac{|\bm s_b(\bm x)^H\bm z_b|^2}{\|\bm s_b(\bm x)\|^2\|\bm z_b\|^2}.
 \label{eq:decision-profile}
\end{equation}
L-BFGS-B uses analytic gradients for at most 12 iterations within radius $0.8$ of initial action $\bm a_0$. Appendix~\ref{app:decisionproofs} derives both gradient and Hessian; only the gradient is used. The receiver checks finiteness and $J_r(\bm a_u)\ge J_r(\bm a_0)-10^{-12}$ before clipping $\bm a_u$ to proposal $\bm a_p\in[-3,3]^2$; that improvement need not survive clipping. It accepts the proposal only if this check and
\begin{equation}
 M(\bm a_p\mid Y)\ge M_\star(Y)-\epsilon_A,
 \qquad \epsilon_A=0.02
 \label{eq:decision-action-budget}
\end{equation}
pass, otherwise retaining $\bm a_0$. For equal prior cell masses, the action-dependent log score is
\begin{equation}
 \begin{split}
 Z(\bm a)={}&\log M(\bm a\mid Y)+\operatorname{LSE}_{ij}\ell_1(\bm x_{ij})\\
 &-\operatorname{LSE}_{ij}\ell_0(\bm x_{ij}),
 \end{split}
 \label{eq:decision-returned-score}
\end{equation}
where LSE denotes log-sum-exp; unequal masses add $\log\pi_{ij}$ to both sums. Accepted proposals update the score by $\log M(\bm a_p)-\log M_\star$, matching the returned action. Retained plateau actions preserve the $10^{-12}$ mass discrepancy, which is not a log-score bound. Total action loss is at most $\epsilon_A+10^{-12}$. Empty delivered sets return $-10^{30}$, a zero placeholder, and rejection. Point-MAP changes action, score, and budget gate together. Algorithm~\ref{alg:receiver} uses strict acceptance with a threshold calibrated for the final score.

\begin{algorithm}[t]
\centering
\caption{Predictive Reacquisition}
\label{alg:receiver}
\small
\setlength{\tabcolsep}{0pt}
\renewcommand{\arraystretch}{1.05}
\begin{tabularx}{\columnwidth}{@{}r@{\hspace{0.6em}}>{\raggedright\arraybackslash}X@{}}
\multicolumn{2}{@{}p{\columnwidth}@{}}{\textbf{Input:} $\{Y_b,Z_b\}$, delivered set $\mathcal A$, assistance $\mathcal I$, independent calibration threshold $\gamma$.}\\
\multicolumn{2}{@{}p{\columnwidth}@{}}{\textbf{Output:} $(\delta,\widehat{\bm x},Z)$. Set $\tau_{\rm n}=10^{-12}$.}\\
\midrule
1 & \textbf{if} $\mathcal A=\varnothing$ \textbf{then return} $(0,\bm0,-10^{30})$.\\
2 & Form shared $q(\bm\eta\mid\mathcal I)$ nodes; reference CIW updates \eqref{eq:pred-reference}.\\
3 & \textbf{for each} $b\in\mathcal A$ \textbf{do}\\
4 & \hangindent=1em\hangafter=0 $p_{0,b}\gets\sum_j\pi_jp_{0,j,b}$, $w_{j,b}^{(0)}\gets\pi_jp_{0,j,b}/p_{0,b}$.\\
5 & \hangindent=1em\hangafter=0 $\mathcal J_b\gets\{0\}\cup\operatorname{Top8}_{j\ne0}w_{j,b}^{(0)}$; evaluate $B_{j,p}$ by \eqref{eq:pred-gamma-bf}--\eqref{eq:pred-tilted-bf} with 16 Gamma nodes.\\
6 & \hangindent=1em\hangafter=0 $\widetilde L_b\gets\sum_{j\in\mathcal J_b}w_{j,b}^{(0)}\sum_pw_pB_{j,p}$, without renormalizing the retained weights.\\
7 & \textbf{end for}; interpolate $\widetilde L_b$ and form $r_{h,bkq}$ using the contamination rule in Section~\ref{sec:predictive}.\\
8 & For each $h\in\{0,1\}$, separately compute\newline
    $\ell_{h,k}\gets\operatorname{LSE}_{q}\!\left(\log\pi_q+\sum_{b\in\mathcal A}\log r_{h,bkq}\right)$.\\
9 & $\Delta_\ell\gets\operatorname{LSE}_k\ell_{1,k}-\operatorname{LSE}_k\ell_{0,k}$,\newline
    $P_k\gets e^{\ell_{1,k}-\operatorname{LSE}_{l}\ell_{1,l}}$ for the equal-prior cells.\\
10 & $M_\star\gets\max_{\bm a\in\mathcal D_{\rm bp}}\bm W(a_d)^TP\bm V(a_f)$ over Proposition~\ref{prop:cell-optimum} breakpoints; choose $\bm a_0$ by the $\tau_{\rm n}$ plateau/mean rule (Section~\ref{sec:action}).\\
11 & $\widehat{\bm x}\gets\bm a_0$, $Z_0\gets\log\max(M_\star,10^{-300})+\Delta_\ell$, $Z\gets Z_0$.\\
12 & Freeze nuisance means at the nearest cell to $\bm a_0$ and observed-data filters; form the normalized profile $J_r$ in \eqref{eq:decision-profile}.\\
13 & Maximize: $\bm a_u\gets\operatorname{L\!{-}\!BFGS\!{-}\!B}_{\le12}(J_r,\nabla J_r;\bm a_0,\bm a_0\pm0.8)$.\\
14 & \textbf{if} $\bm a_u$ is finite and $J_r(\bm a_u)\ge J_r(\bm a_0)-\tau_{\rm n}$ \textbf{then}\\
15 & \hangindent=1em\hangafter=0 $\bm a_p\gets\operatorname{clip}_{\mathcal D}(\bm a_u)$, $M_p\gets M(\bm a_p)$.\\
16 & \hangindent=1em\hangafter=0 \textbf{if} $M_p\ge M_\star-\epsilon_A$, $\epsilon_A=0.02$, \textbf{then}\\
17 & \hangindent=2em\hangafter=0 $\widehat{\bm x}\gets\bm a_p$, $Z\gets Z_0+\log\!\dfrac{\max(M_p,10^{-300})}{\max(M_\star,10^{-300})}$.\\
18 & \hangindent=1em\hangafter=0 \textbf{end if}\\
19 & \textbf{end if}; failed checks retain $(\bm a_0,Z_0)$.\\
20 & $\delta\gets\mathbf1\{Z>\gamma\}$; \textbf{return} $(\delta,\widehat{\bm x},Z)$.\\
\end{tabularx}
\end{algorithm}

For $n_dn_f$ cells and $Q$ shared nodes, fusion costs $\mathcal O(|\mathcal A|n_dn_fQ)$; dense overlap evaluation at $A_d,A_f$ breakpoints costs $\mathcal O(A_dn_dn_f+A_dn_fA_f)$. After spatial filtering, refinement costs $\mathcal O(E_r|\mathcal A|NK_F)$ for $E_r$ function evaluations and $K_F$ Fourier coefficients. Line search permits $E_r>12$. Benchmark job times include shared candidate-field precomputations and do not measure standalone hardware latency.

\section{Theoretical Analysis}\label{sec:analysis}
At fixed $Y$, let $P_Y,\widehat P_Y$ be normalized reference and implemented target posteriors on the same space, with $R_P(\bm a)=P_Y(\mathcal C(\bm a))$ and $\operatorname{TV}(P,Q)=\sup_A|P(A)-Q(A)|$ (half the $L^1$ distance for densities).
\begin{proposition}[Conditional action regret]\label{prop:decision-regret}
If $\operatorname{TV}(P_Y,\widehat P_Y)\le\delta(Y)$ and the implemented action loses at most $\epsilon$ capture mass under $\widehat P_Y$, then
\begin{equation}
 \max_{\bm a}R_P(\bm a)-R_P(\widehat{\bm x})
 \le\min\{1,2\delta(Y)+\epsilon\}.
 \label{eq:decision-regret}
\end{equation}
\end{proposition}
Here $\epsilon=\epsilon_A+10^{-12}$ includes refinement and plateau tolerances. Proposition~\ref{prop:decision-regret} separates posterior error $\delta(Y)$ from computed-posterior action loss. The bound remains conditional on TV error; the action budget certifies neither global TV accuracy nor false-acceptance control under distribution shift, tested empirically in Section~\ref{sec:experiments}.

For positive screening $0\le\widetilde g_Y\le g_Y$ with normalizers $0<\widetilde Z_Y=\int\widetilde g_Y\le Z_Y=\int g_Y$, the full posterior mixes retained and omitted laws with omitted weight $\delta_T=1-\widetilde Z_Y/Z_Y$. Hence truncation TV error is at most $\delta_T$, which omitted null mass cannot bound because discarded target likelihood ratios may be large.

At fixed $\bm x,h$, let $X_b(\bm\eta)=\log p_{h,b}^{\varepsilon}(Y_b\mid\bm x,\bm\eta,\mathcal I)$ and define
\begin{equation}
 D_h(t)=\log\mathbb E_qe^{t\sum_bX_b}-\sum_b\log\mathbb E_qe^{tX_b}.
 \label{eq:decision-dependence-def}
\end{equation}
For normalized tilts $q_{J,t}\propto qe^{t\sum_bX_b}$ and $q_{b,t}\propto qe^{tX_b}$,
\begin{equation}
 \begin{split}
 D_h(1)=\int_0^1(1-t)\bigg\{&\operatorname{Var}_{q_{J,t}}\!\left(\sum_bX_b\right)\\
 &-\sum_b\operatorname{Var}_{q_{b,t}}(X_b)\bigg\}\,dt.
 \end{split}
 \label{eq:decision-dependence-identity}
\end{equation}
This holds for finite positive quadratures; continuous laws require finite exponential moments and differentiation under the integral near $[0,1]$. Neither its sign nor capture improvement is universal. With normalized independent candidate weights $\rho_h^I(\bm x)\propto\pi_x(\bm x)\prod_b\mathbb E_qe^{X_b}$, joint marginalization changes the log evidence ratio by
\begin{equation}
 \log\mathbb E_{\rho_1^I}e^{D_1(1,\bm x)}
 -\log\mathbb E_{\rho_0^I}e^{D_0(1,\bm x)}.
 \label{eq:decision-global-dependence}
\end{equation}
For fixed actions, restrict and renormalize alternative weights on $\mathcal C(\bm a)$; the maximizing action may also change.

Fix the score before calibration. Let $S_1,\ldots,S_{n_c+1}$ be exchangeable null scores including assistance and delivery randomness, with the last score reserved for testing. For $0<\alpha_{\rm FA}<1$, set
\begin{equation}
 k=\lceil(n_c+1)(1-\alpha_{\rm FA})\rceil,\qquad\gamma=S_{(k)},
 \label{eq:decision-rank-threshold}
\end{equation}
using $+\infty$ if $k>n_c$. Strict acceptance gives
\begin{equation}
 \Pr(S_{n_c+1}>\gamma)\le\frac{n_c+1-k}{n_c+1}\le\alpha_{\rm FA},
 \label{eq:decision-rank-bound}
\end{equation}
The first bound is attained for continuous scores; $k>n_c$ gives zero. Appendix~\ref{app:decisionproofs} treats ties. This marginal calibration/test guarantee yields $k=245$ and $12/257$ for $n_c=256$, $\alpha_{\rm FA}=0.05$. Each scenario and seed uses independent matched calibration; distribution shift and likelihood accuracy require separate validation.

\section{Simulation Setup}
\label{sec:setup}

\begin{figure*}[!t]
\centering
\includegraphics[width=.98\textwidth]{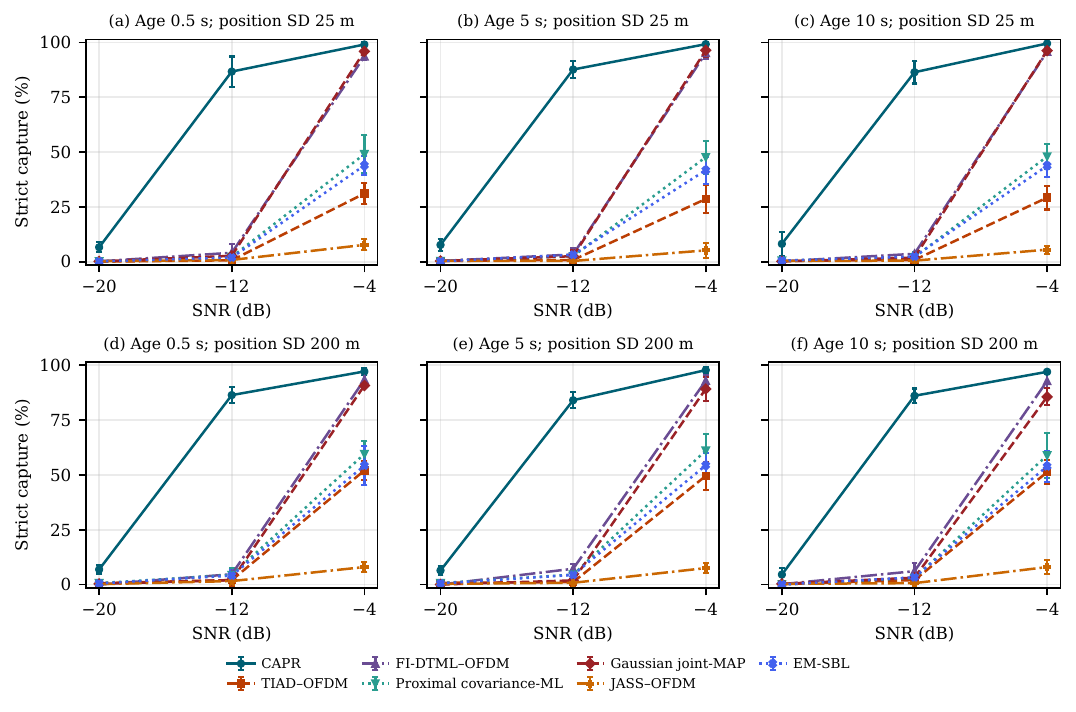}
\caption{Complete crossed replay experiment: all 18 combinations of received SNR, assistance age, and position uncertainty. Each panel fixes age and position uncertainty and retains every evaluated SNR and all six baselines. Error bars are 95\% intervals across five seed repetitions.}
\label{fig:replay-crossed}
\end{figure*}

\begin{figure*}[!t]
\centering
\includegraphics[width=.98\textwidth]{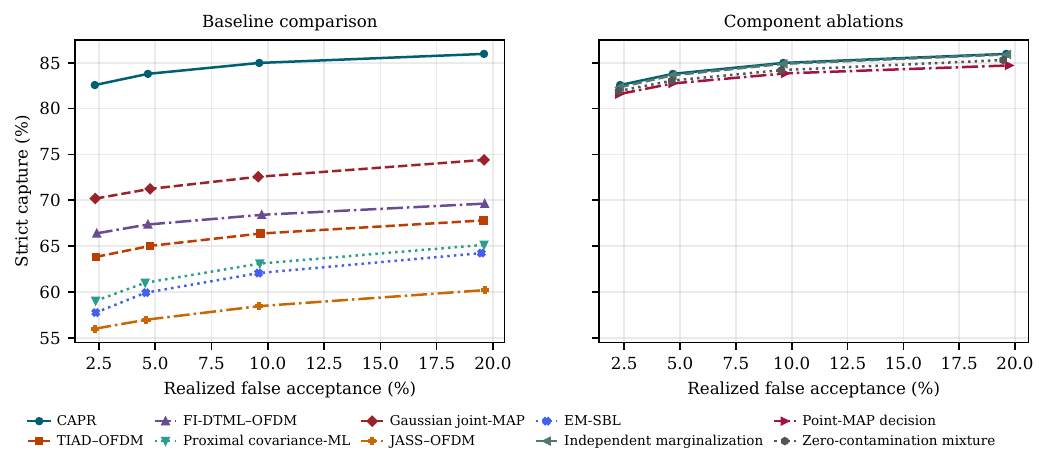}
\caption{Independent-calibration operating points for the unchanged primary scenario set at nominal levels 0.025, 0.05, 0.10, and 0.20. Thresholds use saved calibration scores only. Test scores determine the displayed false acceptance and strict capture; connecting lines identify successive nominal levels.}
\label{fig:operating}
\end{figure*}
\begin{figure*}[!t]
\centering
\includegraphics[width=.98\textwidth]{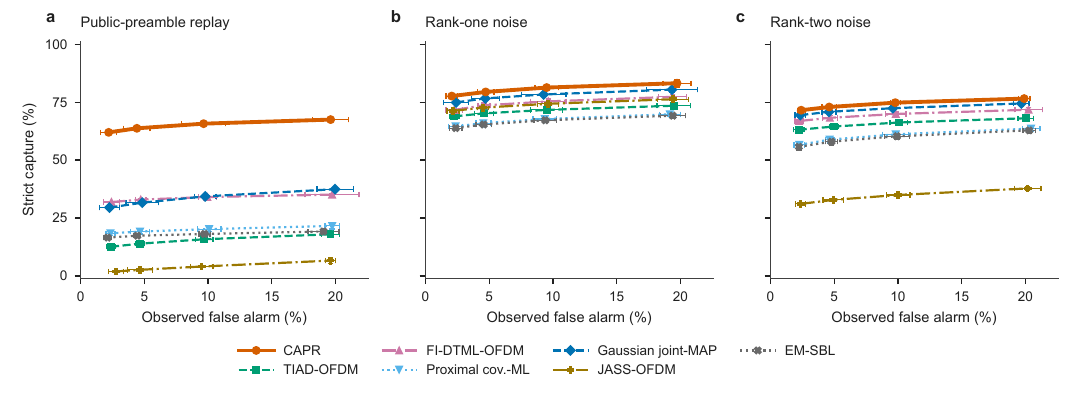}
\caption{Independently calibrated operating points by interference mechanism. Each panel retains the complete 18-condition grid: SNR -20, -12, and -4 dB; clock age 0.5, 5, and 10 s; and position standard deviation 25 and 200 m. Rank-two noise includes a second full-power component. Bars show pointwise 95\% seed intervals on both axes; lines connect the four evaluated nominal levels.}
\label{fig:operating-mechanisms}
\end{figure*}

\begin{figure*}[!t]
\centering
\includegraphics[width=.98\textwidth]{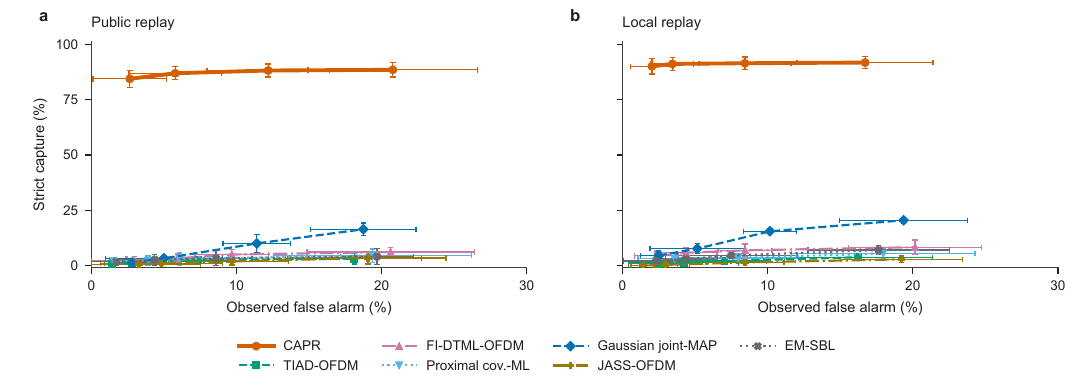}
\caption{Operating points under common replay offsets (a) and node-local replay offsets (b), at SNR -12 dB, clock age 2 s, and position standard deviation 35 m. Each panel retains all seven methods and five seeds. Bars show pointwise 95\% seed intervals on both axes; lines connect independently calibrated nominal levels.}
\label{fig:replay-offsets}
\end{figure*}

The benchmark evaluates \ResultPrimaryCases{} equally weighted primary scenarios and two negative controls across five independent random seeds. Each scenario-seed pair generates 256 target-absent calibration trials and 128 test trials per hypothesis, totaling \ResultCalibration{} calibration and \ResultTests{} test observations. All methods receive identical delivered IQ packets, reference blocks, assistance states, and availability sets. Table~\ref{tab:settings} summarizes the default parameters; individual scenarios override SNR, assistance age, position uncertainty, interference type, multipath profile, quantization depth, reference length, and delivery deadline.

\begin{table}[t]
\centering
\caption{Default physical and numerical settings from the frozen manifest. Scenario overrides are retained; discrete counts are exact.}
\label{tab:settings}
\footnotesize
\setlength{\tabcolsep}{3pt}
\begin{tabular}{ll}
\toprule
Parameter & Value \\
\midrule
Carrier / sample rate & 2.00 GHz / 240 ksample/s \\
BS / UAV / satellite nodes & 2 / 3 / 1 \\
Antennas per node & 4 \\
FFT / cyclic prefix / symbols & 16 / 4 / 4 \\
CFO normalization & \ensuremath{1.50\times10^{3}} Hz \\
Reference samples & 48 \\
IQ bits / backhaul & 12 / 20.0 Mbit/s \\
Deadline / mean queue parameter & 12.0 / 0.0800 ms \\
Grid step / common-state nodes & 0.250 / 64 \\
Retained competitor / Gamma nodes & 8 / 16 \\
Contamination / action-loss budget & 0.0200 / 0.0200 \\
Calibration / tests per hypothesis & 256 / 128 \\
Nominal false acceptance & 5.00\% \\
\bottomrule
\end{tabular}
\end{table}

Strict capture $P_{\rm A}$ requires $H_1$ acceptance and errors within $0.5$ sample and $750$ Hz; threshold-free success $P_{\rm E}$ uses the same box without thresholding, and $P_{\rm FA}$ concerns $H_0$. A physical-time variant fixes delay tolerance at $0.5/240000$ s across sampling rates. Unconditional RMSE includes rejected and empty $H_1$ trials: macro RMSE averages scenario-seed RMSEs, whereas pooled RMSE takes the root after pooling squared errors. Means and 95\% Student-$t$ intervals use five equally weighted seed means and equal scenario weights, with paired intervals from within-seed differences and approximate Gaussianity assumed. Five positive seed differences still give exact two-sided sign-test $p=0.0625$. Independent random stages generate calibration and tests; true parameters serve evaluation only. Audits of paired-input digests, thresholds and summaries give recomputation error below $10^{-12}$, without reconstructing full IQ. Tables rank unrounded capture, estimation and RMSE values within each comparison set; bold and underlining mark the best and second distinct levels, while PFA and intervals are unranked.

Six baselines share this protocol. TIAD-OFDM restricts delay-Doppler search using timing assistance~\cite{li2026tiad}; FI-DTML-OFDM combines bidirectional frequency separation with a graph point estimate~\cite{jin2024fidtml}, omitting its full timestamp tracker. Proximal covariance-ML uses a log-determinant/trace objective with a sum-minus-maximum penalty~\cite{li2023asynccov}. EM-SBL uses Gaussian coefficient posteriors and EM variance updates~\cite{wipf2004sbl,wipf2007simultaneous}, distinct from structured EM-MVSP~\cite{zuo2023ofdm}. Both sparse methods retain 14 of 49 candidates and run 20 iterations. Gaussian joint-MAP jointly profiles synchronization and Gaussian assistance states for 16 iterations~\cite{teng2026blacs}, without BLACS messages. JASS-OFDM combines rank-one projection with a CFO bank~\cite{marti2025jass}; public replay exceeds its secret-sequence premise. The five two-stage baselines use at most 12 frozen-nuisance refinement iterations and retain grid detection scores. CAPR uses the same refinement cap with posterior nuisance expectations and the mass gate~\eqref{eq:decision-action-budget}. This compares fixed-iteration accuracy; differing inference stages and line searches need not have equal runtime.

\section{Numerical Experiments}
\label{sec:experiments}
\label{sec:replay-results}
\begin{table*}[!t]
\centering
\caption{Primary performance over all 150 primary scenarios and five seeds. RMSEs average scenario RMSEs.}
\label{tab:main-results}
\footnotesize
\setlength{\tabcolsep}{3pt}
\begin{tabular}{p{.28\textwidth}rrrrrr}
\toprule
Implementation & $P_{\rm A}$ (\%) $\uparrow$ & 95\% seed CI & $P_{\rm FA}$ (\%) & $P_{\rm E}$ (\%) $\uparrow$ & Delay (sample) $\downarrow$ & CFO (Hz) $\downarrow$ \\
\midrule
\CAPR & $\mathbf{83.8}$ & [83.5, 84.1] & 4.68 & $\mathbf{87.8}$ & $\mathbf{0.430}$ & $\mathbf{611}$ \\
Gaussian joint-MAP~\cite{teng2026blacs} & $\underline{71.2}$ & [71.0, 71.5] & 4.77 & $\underline{86.2}$ & $\underline{0.450}$ & $\underline{659}$ \\
FI-DTML--OFDM~\cite{jin2024fidtml} & 67.4 & [67.1, 67.6] & 4.68 & 73.3 & 0.717 & 945 \\
TIAD--OFDM~\cite{li2026tiad} & 65.0 & [64.8, 65.2] & 4.78 & 72.2 & 0.907 & \ensuremath{1.13\times10^{3}} \\
Proximal covariance-ML~\cite{li2023asynccov} & 61.0 & [60.7, 61.3] & 4.55 & 70.7 & 1.02 & \ensuremath{1.30\times10^{3}} \\
EM-SBL~\cite{wipf2004sbl,wipf2007simultaneous} & 59.9 & [59.6, 60.2] & 4.59 & 70.0 & 1.04 & \ensuremath{1.38\times10^{3}} \\
JASS--OFDM~\cite{marti2025jass} & 57.0 & [56.8, 57.2] & 4.61 & 66.8 & 1.02 & \ensuremath{1.33\times10^{3}} \\
\bottomrule
\end{tabular}
\end{table*}

\begin{table*}[!t]
\centering
\caption{Strict capture (\%, $\uparrow$) by manifest interference kind. Each scenario and then each seed has equal weight.}
\label{tab:family-results}
\footnotesize
\setlength{\tabcolsep}{3pt}
\begin{tabular}{lrrrrrrrr}
\toprule
Interference kind & Cases & \CAPR{} & TIAD & FI-DTML & Prox-ML & Joint-MAP & JASS & EM-SBL \\
\midrule
Noise only & 1 & $\underline{99.5}$ & $\underline{99.5}$ & 95.2 & 99.4 & $\mathbf{99.8}$ & $\underline{99.5}$ & 99.4 \\
Barrage & 89 & $\mathbf{88.6}$ & 83.1 & 81.5 & 77.6 & $\underline{87.4}$ & 84.8 & 76.5 \\
Replay & 31 & $\mathbf{72.8}$ & 8.46 & $\underline{20.6}$ & 12.2 & 19.6 & 2.09 & 11.0 \\
Rank-two & 26 & $\mathbf{78.5}$ & 69.2 & 73.9 & 61.0 & $\underline{76.6}$ & 24.6 & 59.7 \\
Tone & 1 & $\underline{98.3}$ & 97.2 & 94.2 & 90.9 & $\mathbf{99.5}$ & 97.3 & 89.4 \\
Chirp & 1 & $\underline{98.1}$ & 97.5 & 95.0 & 93.0 & $\mathbf{98.3}$ & 97.8 & 91.9 \\
Local replay & 1 & $\mathbf{91.1}$ & 1.25 & 5.78 & 3.13 & $\underline{7.66}$ & 0.313 & 2.97 \\
\midrule
All primary scenarios & 150 & $\mathbf{83.8}$ & 65.0 & 67.4 & 61.0 & $\underline{71.2}$ & 57.0 & 59.9 \\
\bottomrule
\end{tabular}
\end{table*}

Across all \ResultPrimaryCases{} primary scenarios (Table~\ref{tab:main-results}), CAPR achieves \ResultGfourCapture\% strict capture at \ResultGfourPFA\% realized $P_{\mathrm{FA}}$ and \ResultGfourEstimate\% threshold-free estimation accuracy. Compared with the strongest baseline, Gaussian joint-MAP (71.2\%), CAPR yields a paired capture gain of $\ResultBestGain$ pp ($95\%~\mathrm{CI}: [\ResultBestGainLow, \ResultBestGainHigh]~\mathrm{pp}$, Holm-adjusted paired-$t$ $p < 10^{-6}$ across all six baselines), recording \ResultBestWins{} scenario wins, \ResultBestTies{} ties, and \ResultBestLosses{} losses. Macro delay and CFO RMSEs reach $\ResultGfourDelay$~sample ($\ResultMacroDelayUs\,\mu\mathrm{s}$) and $\ResultGfourCFO$ Hz, improving upon Gaussian joint-MAP by \ResultDelayReduction\% and \ResultCFOReduction\%, respectively (pooled RMSEs are $\ResultPooledDelay$~sample / $\ResultPooledDelayUs\,\mu\mathrm{s}$ and $\ResultPooledCFO$ Hz, with \ResultPhysicalCapture\% fixed-time capture). Event decomposition shows that CAPR improves threshold-free estimation $P_E$ by $\ResultEstimateGain$ pp while reducing valid-but-rejected estimates by $\ResultRejectedCorrectReduction$ pp. Incorrect $H_1$ acceptance also increases by $\ResultAcceptedIncorrectIncrease$ pp; this decomposition does not isolate a single algorithmic component.

As shown in Table~\ref{tab:family-results}, the overall improvement is driven primarily by resilience to public-preamble replay. Across all \ResultReplayCases{} common (shared-offset) and node-local (independent-offset) replay scenarios, CAPR achieves \ResultReplayCapture\% strict capture versus \ResultReplayGaussian\% for Gaussian joint-MAP, accounting for \ResultReplayShare\% of the aggregate benchmark gain (while providing a $\ResultOtherGain$ pp gain across the remaining \ResultOtherCases{} scenarios). CAPR explicitly represents temporal preamble competition under both hypotheses, whereas Gaussian joint-MAP models interference through spatial reference covariance; the gain is consistent with mitigating replay-induced target ambiguity. Fig.~\ref{fig:replay-crossed} illustrates the 18 crossed SNR-age-position replay scenarios, and Table~\ref{tab:replay-contexts} reports the remaining 14 replay configurations spanning 3D geometry, 64-subcarrier waveforms, tight deadlines, clock correlations, and reference visibility levels. At $-20~\mathrm{dB}$ SNR and $200~\mathrm{m}$ position uncertainty, capture drops to \ResultSevereFiveCapture\% and \ResultSevereTenCapture\% at ages $5~\mathrm{s}$ and $10~\mathrm{s}$ ($P_E$ of \ResultSevereFiveEstimate\% and \ResultSevereTenEstimate\%), showing that relative gains coexist with low absolute reacquisition probability.

\begin{table*}[!t]
\centering
\caption{All replay contexts outside the crossed SNR--age--position grid: strict capture (\%, $\uparrow$) averaged over the five seeds. These fourteen scenarios and the eighteen crossed scenarios exhaust the replay focus.}
\label{tab:replay-contexts}
\footnotesize
\setlength{\tabcolsep}{3.2pt}
\begin{tabular}{p{.30\textwidth}rrrrrrr}
\toprule
Physical context & \CAPR{} & TIAD & FI-DTML & Prox-ML & Joint-MAP & JASS & EM-SBL \\
\midrule
\multicolumn{8}{l}{\textit{Baseline geometry}} \\
Public replay & $\mathbf{86.9}$ & 1.09 & $\underline{3.28}$ & 2.19 & $\underline{3.28}$ & 0.938 & 2.66 \\
Local replay & $\mathbf{91.1}$ & 1.25 & 5.78 & 3.13 & $\underline{7.66}$ & 0.313 & 2.97 \\
\addlinespace[2pt]
\multicolumn{8}{l}{\textit{3D geometry and clock age}} \\
Age 0.5 s, position SD 35 m & $\mathbf{77.2}$ & 0.156 & 0.938 & $\underline{1.72}$ & 1.09 & 0.469 & 1.25 \\
Age 0.5 s, position SD 200 m & $\mathbf{76.6}$ & 1.41 & $\underline{4.06}$ & 3.75 & 2.03 & 1.25 & 2.81 \\
Age 8 s, position SD 35 m & $\mathbf{82.3}$ & 0.781 & $\underline{2.03}$ & 1.09 & 0.938 & 1.09 & 1.25 \\
Age 8 s, position SD 200 m & $\mathbf{73.4}$ & 1.25 & $\underline{3.28}$ & 3.13 & 2.81 & 1.09 & 2.66 \\
\addlinespace[2pt]
\multicolumn{8}{l}{\textit{3D OFDM and delivery}} \\
64 subcarriers, 100 m position SD & $\mathbf{98.1}$ & 2.19 & 10.2 & 5.47 & $\underline{18.1}$ & 7.19 & 4.69 \\
Delivery deadline 4 ms & $\mathbf{85.8}$ & 0.781 & $\underline{5.16}$ & 2.34 & 4.38 & 0.938 & 2.66 \\
\addlinespace[2pt]
\multicolumn{8}{l}{\textit{3D clock correlation}} \\
Clock correlation $\rho=0$ & $\mathbf{86.3}$ & 1.09 & $\underline{3.91}$ & 2.81 & 0.469 & 0.938 & 2.97 \\
Clock correlation $\rho=0.8$ & $\mathbf{89.2}$ & 0.938 & $\underline{6.25}$ & 2.03 & 1.41 & 0.781 & 2.19 \\
Clock correlation $\rho=0.95$ & $\mathbf{87.3}$ & 0.469 & 3.44 & $\underline{3.75}$ & 1.41 & 0.938 & 2.81 \\
\addlinespace[2pt]
\multicolumn{8}{l}{\textit{3D reference visibility}} \\
Reference visibility $0$ & $\mathbf{91.1}$ & 0.156 & 0.156 & 0.156 & 0.00 & $\underline{0.313}$ & 0.156 \\
Reference visibility $0.5$ & $\mathbf{86.6}$ & 0.156 & 0.313 & 0.156 & 0.313 & $\underline{0.625}$ & $\underline{0.625}$ \\
Reference visibility $1$ & $\mathbf{88.6}$ & 0.625 & 2.50 & $\underline{3.28}$ & 1.09 & 0.469 & 3.13 \\
\bottomrule
\end{tabular}
\par\smallskip
\end{table*}

\subsection{Operating Characteristics and Component Ablations}
Fig.~\ref{fig:operating} plots realized strict capture against false acceptance across nominal calibration levels $\alpha_{\mathrm{FA}} \in \{0.025, 0.05, 0.10, 0.20\}$ using thresholds determined strictly from saved calibration scores. Figs.~\ref{fig:operating-mechanisms} and~\ref{fig:replay-offsets} further stratify these operating curves by interference mechanism and common versus node-local replay offsets, showing capture gains in these configurations. Replay gains over Gaussian joint-MAP exceed both noise-mechanism gains at every evaluated level. Each scenario and seed is calibrated separately; actual PFA differs across methods. Mechanism plots retain all 18 matched conditions; rank-two noise adds a full-power component. Offset examples use $-12$ dB, 2-s age and 35-m position uncertainty. Intervals are clipped to $[0,100]\%$, and lines connect evaluated points only.

\begin{table*}[!t]
\centering
\caption{Component ablations on identical observations. Point-MAP changes the action, score, and proposal-budget gate together.}
\label{tab:ablations}
\footnotesize
\setlength{\tabcolsep}{3pt}
\begin{tabular}{p{.28\textwidth}rrrrrr}
\toprule
Implementation & $P_{\rm A}$ (\%) $\uparrow$ & 95\% seed CI & $P_{\rm FA}$ (\%) & $P_{\rm E}$ (\%) $\uparrow$ & Delay (sample) $\downarrow$ & CFO (Hz) $\downarrow$ \\
\midrule
\CAPR & $\mathbf{83.8}$ & [83.5, 84.1] & 4.68 & $\underline{87.8}$ & 0.430 & 611 \\
Independent marginalization & $\underline{83.6}$ & [83.3, 84.0] & 4.66 & 87.8 & $\underline{0.430}$ & $\underline{609}$ \\
Point-MAP decision & 82.7 & [82.5, 82.9] & 4.66 & 86.3 & 0.451 & 647 \\
Zero-contamination mixture & 83.0 & [82.7, 83.4] & 4.64 & $\mathbf{87.9}$ & $\mathbf{0.430}$ & $\mathbf{609}$ \\
\bottomrule
\end{tabular}
\end{table*}

Table~\ref{tab:ablations} evaluates three component ablations on identical observations: 1) \emph{Independent marginalization}, which factors shared-state expectations and proposal nuisance weights across nodes; 2) \emph{Point-MAP decision}, which changes the action, score and refinement gate together; and 3) \emph{Zero-contamination mixture} ($\varepsilon = 0$). Removing these components degrades strict capture by $\ResultIndependentGain$ pp ($\mathrm{CI}: [\ResultIndependentGainLow, \ResultIndependentGainHigh]$), $\ResultPeakGain$ pp ($[\ResultPeakLow, \ResultPeakHigh]$), and $\ResultZeroGain$ pp ($[\ResultZeroLow, \ResultZeroHigh]$), respectively, supporting their combined contribution. Independent marginalization changes estimation success little; zero contamination slightly improves unconditional RMSE, and correlation-conditioned gains can change sign across interference types.

\subsection{Operational Boundaries and Scope Discussions}
\label{sec:experimental-boundaries}
We next examine four regimes that define CAPR's operational boundaries. In the reported correlation-conditioned comparison, across prior clock correlations of $0$, $0.8$, and $0.95$, replay capture gains are $\ResultCorrReplayZero$, $\ResultCorrReplayEight$, and $\ResultCorrReplayNineFive$ pp (where $\rho = 0$ still retains shared geometry and timing-graph conditioning). In diffuse non-line-of-sight channels, CAPR attains \ResultDiffuseCapture\% capture versus \ResultDiffuseGaussian\% for Gaussian joint-MAP and \ResultDiffuseBest\% for JASS-OFDM, consistent with angular scattering departing from the target model's rank-one steering structure. Under clock-prior mismatch, FI-DTML-OFDM (\ResultClockOODBest\%) exceeds CAPR (\ResultClockOODCapture\%), reflecting sensitivity to misspecified prior scales. All such adverse scenarios are retained within the primary \ResultPrimaryCases{}-scenario averages.
Negative controls illustrate the identifiability limit of public-preamble synchronization. When the target and competitor obey identical observable laws, CAPR yields \ResultIdentityDetection\% $\mathcal{H}_1$ acceptance at \ResultIdentityPFA\% $P_{\mathrm{FA}}$ despite \ResultIdentityEstimate\% estimation accuracy; under near-zero-offset replay, capture falls to \ResultNearReplayCapture\% (\ResultNearReplayDetection\% detection at \ResultNearReplayPFA\% $P_{\mathrm{FA}}$). Accurate synchronization cannot identify sources with identical observable laws; near replay also degrades estimation. Public-preamble synchronization therefore cannot substitute for transmitter authentication.
Diagnostics across all \ResultValidActions{} nonempty test trials confirm that every returned action obeys the posterior mass-loss budget $\epsilon_A = 0.0200$ (with the \ResultEmptyActions{} empty observations retained in unconditional metrics). The displayed maximum loss is $\ResultMaximumGap$. This verifies the working-posterior gate, not solver convergence or physical-model regret (Proposition~\ref{prop:decision-regret}). Screening trades accuracy for computation; its maximum omitted null weight, $\ResultMaximumOmitted$, bounds neither omitted alternative mass nor posterior TV error. Algorithm~\ref{alg:receiver} couples the returned score to action mass up to the stated plateau tolerance.
The $1~\mathrm{ms}$ deadline stress test yields zero delivered packets and 0\% capture due to propagation, sampling, serialization, and queueing floors. Sweeps over antenna ports and reference lengths exhibit nonmonotone variations under distinct observations and coupled power/delivery conditions, precluding an isolated causal interpretation. The deadline excludes receiver CPU time. Hardware profiling, payload bit-error rate (BER), throughput and dynamic timing-graph scheduling remain natural extensions for cross-layer implementations.

\FloatBarrier
\section{Conclusion}
\label{sec:conclusion}
This paper presented CAPR, a predictive Bayesian receiver for OFDM burst reacquisition across clock-asynchronous SAGINs in the presence of competing public preambles. By analytically marginalizing spatial covariance and target complex amplitude under both hypotheses and fusing evidence across shared geometry and clock states, CAPR improves burst recovery under coherent replay in the evaluated regimes. Coupling exact finite-cell capture-mass optimization with mass-budgeted continuous refinement aligns the detection statistic directly with downstream synchronization tolerances. Across \ResultPrimaryCases{} primary scenarios, CAPR achieved \ResultGfourCapture\% strict capture at \ResultGfourPFA\% false acceptance, improving capture over Gaussian joint-MAP by \ResultBestGain{} percentage points overall and by \ResultReplayGain{} percentage points across all \ResultReplayCases{} replay scenarios. Future work will explore adaptive reference acquisition, online calibration under distribution shift, and hardware-accelerated implementation.

\appendices
\section{Predictive Likelihood Derivations}
\label{app:predictive}
Column vectorization gives
$\operatorname{vec}(E)^H(B\otimes C)\operatorname{vec}(E)=\operatorname{tr}(CEB^TE^H)$,
hence $V^{-T}$ in \eqref{eq:pred-matrix-normal}. On Hermitian measure, $X=\Sigma^{-1}$ and $W=B^{1/2}XB^{1/2}$ give $\mathrm d\Sigma=|X|^{-2M}\mathrm dX$ and $\mathrm dX=|B|^{-M}\mathrm dW$. For $B\succ0$, $a>M-1$,
\begingroup\interdisplaylinepenalty=10000
\begin{align}
 &\int_{\Sigma\succ0}|\Sigma|^{-(a+M)}
 e^{-\operatorname{tr}(B\Sigma^{-1})}\,\mathrm d\Sigma\nonumber\\
 &=|B|^{-a}\int_{W\succ0}|W|^{a-M}e^{-\operatorname{tr}W}\,\mathrm dW
 =|B|^{-a}\Gamma_M(a).
 \label{eq:pred-app-ciw-integral}
\end{align}
\endgroup
Reference and capture conditioning add $(ZZ^H,T)$ and $(EV^{-T}E^H,N)$ to scale and shape. Substituting $B=\Psi+EV^{-T}E^H$, $a=K$ yields \eqref{eq:pred-matrix-student}. The determinant $|V\otimes\Sigma|=|V|^M|\Sigma|^N$ normalizes the Gaussian; Tonelli normalizes its proper covariance mixture.

For $F_j=C_jD_j^{1/2}$ and $H_j=I_{r_j}-F_j^HR^{-1}F_j$, the determinant lemma gives $|R_j|=|R|\,|H_j|$, while Woodbury gives
\begin{equation}
 R_j^{-1}=R^{-1}+R^{-1}F_jH_j^{-1}F_j^HR^{-1}.
 \label{eq:pred-app-woodbury}
\end{equation}
Here $R_j=R-F_jF_j^H\succ0$ ensures $H_j\succ0$; zero eigenvalues of $D_j$ may be omitted.

Whitening by $\widetilde Y=YV_j^{-T/2}$, set $\bm t=V_j^{-1/2}\bm s/\sqrt{e_j}$ and unitary $Q=[\bm t^*,Q_\perp]$. Then $\widetilde YQ=[\bm c_j,Y_\perp]$ and $\bm t^TQ=[1,0,\ldots,0]$ confine target mean $\alpha\sqrt{e_j}\bm a_{\rm s}$ to $\bm c_j$. The remaining columns yield $\operatorname{CIW}_M(S_j,K-1)$ with $S_j=\Psi+Y_\perp Y_\perp^H$. Covariance integration gives
\begingroup\interdisplaylinepenalty=10000
\begin{align}
 p(\bm c_j\mid Y_\perp,H_0)
 &=\frac{\Gamma_M(K)|S_j|^{K-1}}
 {\pi^M\Gamma_M(K-1)|S_j+\bm c_j\bm c_j^H|^K}\nonumber\\
 &=\frac{\Gamma(K)(1+q_j)^{-K}}{\pi^M\Gamma(K-M)|S_j|},
 \label{eq:pred-app-projected-student}
\end{align}
\endgroup
using $\Gamma_M(K)/\Gamma_M(K-1)=\Gamma(K)/\Gamma(K-M)$. Mixing $\mathcal{CN}(0,S_j/z)$ with $\operatorname{Gamma}(K-M,1)$ gives kernel $z^{K-1}e^{-(1+q_j)z}$ and divisor $\pi^M|S_j|\Gamma(K-M)$. Thus
\begin{equation}
 \int_0^\infty z^{K-1}e^{-(1+q_j)z}\,\mathrm dz
 =\frac{\Gamma(K)}{(1+q_j)^K}
 \label{eq:pred-app-gamma-mixture}
\end{equation}
recovers \eqref{eq:pred-app-projected-student} and null posterior $\operatorname{Gamma}(K,1+q_j)$.

Fixed $\alpha$ replaces $\bm c_j$ by $\bm c_j-\alpha\sqrt{e_j}\bm a_{\rm s}$ without changing $S_j$. Its prior is independent of covariance and target-free columns, permitting amplitude integration first by Tonelli. Set $b_\alpha=\sqrt{e_j}z\bm a_{\rm s}^HS_j^{-1}\bm c_j$, $A_\alpha=\gamma_p^{-1}+ze_ju_j>0$, and $\rho=\gamma_pe_j$. Completing the square gives
\begingroup\interdisplaylinepenalty=10000
\begin{align}
 &\frac1{\pi\gamma_p}\int_{\mathbb C}
 e^{-A_\alpha|\alpha|^2+2\operatorname{Re}(\alpha^*b_\alpha)}\,\mathrm d^2\alpha\nonumber\\
 &\quad=\frac{e^{|b_\alpha|^2/A_\alpha}}{\pi\gamma_p}
 \int_{\mathbb C}e^{-A_\alpha|\alpha-b_\alpha/A_\alpha|^2}\,\mathrm d^2\alpha\nonumber\\
 &\quad=\frac{e^{|b_\alpha|^2/A_\alpha}}{\gamma_pA_\alpha}
 =\frac{\exp\{\rho v_jz^2/(1+\rho u_jz)\}}{1+\rho u_jz}.
 \label{eq:pred-app-alpha-integral}
\end{align}
\endgroup
For coordinate densities $f_0,f_1$ conditional on $(z,Y_\perp)$, with $f_1$ averaged over $\alpha$, and original Gamma density $\pi_z$, the identity
\begin{equation}
 \frac{\int f_1\pi_z\,\mathrm dz}{\int f_0\pi_z\,\mathrm dz}
 =\int\frac{f_1}{f_0}\,p(z\mid\bm c_j,H_0)\,\mathrm dz,
 \label{eq:pred-app-null-posterior-integral}
\end{equation}
proves \eqref{eq:pred-gamma-bf}. Cauchy--Schwarz gives $\theta=v_j/u_j\le q_j$. With $r=1+q_j$, the exponent is $\theta z-\theta z/(1+\rho u_jz)$; the identity
$e^{\theta z}g_{K,r}(z)=[r/(r-\theta)]^K g_{K,r-\theta}(z)$
and $r-\theta\ge1$ prove \eqref{eq:pred-tilted-bf}.

Each $p_{T,j}=p_{0,j}\sum_pw_pB_{T\mid j,p}$ is normalized. Multiplying \eqref{eq:pred-null-reweight} by $p_0$ recovers the prior mixture and its null-posterior weights. For measurable retention $I_j(Y)=\mathbf1\{j\in\mathcal J(Y)\}$,
\begin{equation}
 \int\widetilde p_T\,\mathrm dY
 =\sum_j\pi_j\int I_j(Y)p_{T,j}(Y)\,\mathrm dY\le1.
 \label{eq:pred-app-subdensity}
\end{equation}
holds for exact component integrals and positive unrenormalized screening. Data-independent convex interpolation and contamination preserve complete densities. Integrating \eqref{eq:pred-network-marginals} over $Y$ and normalized assistance/candidate priors normalizes each hypothesis. Canceling $\prod_bp_{0,b}(Y_b)$ retains the contaminated null's candidate integral; screened inputs yield only subdensities.

Finite-precision safeguards floor $e_j,u_j$ at $10^{-10},10^{-15}$, clip $\bm c_j^HR_j^{-1}\bm c_j$ to $[0,1-10^{-10}]$, enforce $\theta\le r-1$, and floor small-matrix eigenvalues at $10^{-12}$ for nonpositive computed determinants. These safeguards and 16-node generalized-Laguerre quadrature contribute to $\delta(Y)$ in Proposition~\ref{prop:decision-regret}; they provide no uniform posterior-error bound.

\section{Proofs of Main Results}\label{app:decisionproofs}
\subsection{Task Optimality and Refinement Derivatives}
Condition on $\mathcal I$. Compactness of $\mathcal D$ and $L^1$ translation continuity of integrable densities ensure a maximum of $G$, with measurable selection as in Theorem~\ref{thm:capture-optimal}. For $D=(\delta,\widehat{\bm x})$ and $\lambda\ge0$,
\begin{align}
 P_A(D)-\lambda P_{\rm FA}(D)
 &=\int\delta(Y)[G(Y,\widehat{\bm x})-\lambda p_0(Y)]\,dY\nonumber\\
 &\le\int[G_\star(Y)-\lambda p_0(Y)]_+\,dY.
 \label{eq:app-decision-lagrange}
\end{align}
Equality selects a maximizing action and accepts $T=G_\star/p_0>\lambda$ on $p_0>0$, or freely where $p_0=0<G_\star$. Choose $\lambda$ whose strict/inclusive null tails bracket $\alpha_{\rm FA}$, randomizing ties with $\rho_\lambda=[\alpha_{\rm FA}-P_0(T>\lambda)]/P_0(T=\lambda)$ when the denominator is positive. If accepting all useful observations leaves slack, set $\lambda=0$. Subtracting this rule $D_\lambda$'s equality from any feasible rule's inequality yields
\begin{equation}
 P_A(D)-P_A(D_\lambda)
 \le\lambda[P_{\rm FA}(D)-P_{\rm FA}(D_\lambda)]\le0.
 \label{eq:app-decision-constrained}
\end{equation}
proving Theorem~\ref{thm:capture-optimal}; randomized actions cannot exceed the pointwise maximum.

Between shifted cell edges and action boundaries, \eqref{eq:decision-cell-mass} is bilinear. At relative coordinates $s,t\in[0,1]$, its corner weights are $(1-s)(1-t)$, $s(1-t)$, $(1-s)t$, and $st$. This convex combination proves Proposition~\ref{prop:cell-optimum}, including partial overlaps and nonuniform cells; continuity covers degeneracy. Accepted proposals satisfy \eqref{eq:decision-action-budget}; rejection retains a plateau action losing at most $10^{-12}$. Both branches therefore lose at most $\epsilon_A+10^{-12}$. The accepted score increment $\log M(\bm a_p)-\log M_\star$ gives the returned-action mass in \eqref{eq:decision-returned-score}.

For one projected node, let $E=\bm s^H\bm s>0$, $V=\bm z^H\bm z>0$, $c=\bm s^H\bm z$, and $A=|c|^2$. Completing the square gives $\|\bm z-\alpha\bm s\|^2=V-A/E+E|\alpha-c/E|^2$ and profile $q=A/(EV)$. For $i,j\in\{d,f\}$, define $c_i=\bm s_i^H\bm z$, $c_{ij}=\bm s_{ij}^H\bm z$, $A_i=2\operatorname{Re}(c^*c_i)$, $A_{ij}=2\operatorname{Re}(c_i^*c_j+c^*c_{ij})$, $E_i=2\operatorname{Re}(\bm s^H\bm s_i)$, and $E_{ij}=2\operatorname{Re}(\bm s_i^H\bm s_j+\bm s^H\bm s_{ij})$. Quotient differentiation gives
\begin{align}
 q_i&=\frac{A_i}{EV}-\frac{AE_i}{E^2V},\nonumber\\
 q_{ij}&=\frac{A_{ij}}{EV}-\frac{A_iE_j+A_jE_i+AE_{ij}}{E^2V}
 +\frac{2AE_iE_j}{E^3V}.
 \label{eq:app-decision-profile-hessian}
\end{align}
Averaging gives $J_r$. Frozen filters and assistance corrections fix $\bm z$ and safeguarded $V=\max(\|\bm z\|^2,10^{-12})$; template-energy derivatives remain. For finite Fourier interpolant $p$, set $s_n=p(v_n)e^{\mathrm i\omega_nf}$, $v_n=n-d-\epsilon[n-(N-1)/2]$, $\omega_n=2\pi\Delta\nu t_n$, and fixed $\epsilon=-\overline f_b\Delta\nu/f_c$. Then $s_d=-p'(v)e^{\mathrm i\omega f}$, $s_f=\mathrm i\omega s$, $s_{dd}=p''(v)e^{\mathrm i\omega f}$, $s_{ff}=-\omega^2s$, and $s_{df}=\mathrm i\omega s_d$. Finite iterations do not guarantee global profile optimality.

\subsection{Approximation Regret and Calibration}
For $\bm a_P\in\arg\max R_P$ and implemented action gap $\epsilon$,
\begin{align}
 R_P(\bm a_P)-R_P(\widehat{\bm x})
 \le{}&|R_P(\bm a_P)-R_{\widehat P}(\bm a_P)|+\epsilon\nonumber\\
 &+|R_{\widehat P}(\widehat{\bm x})-R_P(\widehat{\bm x})|.
 \label{eq:app-decision-regret-steps}
\end{align}
Each exterior term is at most $\delta(Y)$; regret is also at most one, proving \eqref{eq:decision-regret}. For the complete working posterior $P_Y^{\rm w}$, $\operatorname{TV}(P_Y,\widehat P_Y)\le\operatorname{TV}(P_Y,P_Y^{\rm w})+\operatorname{TV}(P_Y^{\rm w},\widehat P_Y)$. All posteriors condition on the same observation; neither the action budget nor omitted null weight bounds these discrepancies.

Under the stated differentiation conditions, $F_X(t)=\log\mathbb E_qe^{tX}$ has $F_X'(t)=\mathbb E_{q_t}X$, $F_X''(t)=\operatorname{Var}_{q_t}X$, and normalized $q_t\propto qe^{tX}$. Applying this to $\sum_bX_b$ and each $X_b$, then integrating twice with $D_h(0)=D_h'(0)=0$, proves \eqref{eq:decision-dependence-identity}. Averaging $e^{D_h(1,\bm x)}$ against $\rho_h^I$ and taking alternative-minus-null logarithms gives \eqref{eq:decision-global-dependence}; restricting and renormalizing $\rho_1^I$ gives the fixed-action case.

For distinct exchangeable null scores, test rank $R$ is uniform on $\{1,\ldots,n_c+1\}$, so strict acceptance $R\ge k+1$ has probability $(n_c+1-k)/(n_c+1)$. Independent uniform tie-breakers preserve uniform ranks; strict exceedance without tie-breaking is a subset of this event. Thus \eqref{eq:decision-rank-bound} holds, with zero acceptance for $k>n_c$. The score is fixed independently of calibration; the guarantee averages over calibration and testing.

\bibliographystyle{IEEEtran}

\end{document}